\documentclass[letterpaper]{article}

\usepackage[T1]{fontenc}

\usepackage{geometry}
\usepackage{setspace}

\usepackage[style = chem-acs]{biblatex}
\let\cite\supercite 

\usepackage{graphicx}
\usepackage{float}
\newfloat{scheme}{htbp}{los}
\floatname{scheme}{Scheme}
\floatname{chart}{Chart}
\newfloat{graph}{htbp}{loh}

\usepackage{chemformula}
\usepackage[version = 4]{mhchem}
\usepackage{caption}

\usepackage{amsmath}
\usepackage{amssymb}

\usepackage[capitalise, noabbrev]{cleveref}

\usepackage{graphicx}
\usepackage{float}
\usepackage{upgreek}

\usepackage{pifont}
\usepackage{booktabs}

\usepackage{rotating} 
\usepackage[table]{xcolor}

\newcommand{\cmark}{\ding{51}}
\newcommand{\xmark}{\ding{55}}

\makeatletter
\newcommand*{\rom}[1]{\expandafter\@slowromancap\romannumeral #1@}
\makeatother

\newcommand{\vb}[1]{\mathbf{#1}}
\newcommand{\ket}[1]{|#1\rangle}

\usepackage{authblk}
\author[1]{Johannes Bürger*}
\affil[1]{CNR Nanotec, Institute of Nanotechnology, Via Monteroni, 73100 Lecce, Italy}
\author[2]{Erika Cortese}
\affil[2]{School of Physics and Astronomy, University of Southampton, Southampton SO17 1BJ, UK}
\author[3]{Simone Di Muzio}
\affil[3]{Istituto di Fotonica e Nanotecnologie—Consiglio Nazionale delle Ricerche (CNR), Piazza Leonardo da Vinci 32, Milano, Italy}
\author[1]{Daniele Sanvitto}
\author[3]{Simone De Liberato}
\author[1]{Dario Ballarini}

\title{Exciton-Tunable Phase Control and Superstrong Coupling Through Multimode Polariton Engineering in Planar Waveguides}

\date{*Email: johannespascalburger@cnr.it}

\begin{document}

\maketitle

\begin{abstract}
Strong light–matter coupling in optical waveguides provides a versatile platform for engineering hybrid polaritonic modes and their dispersion. Here we investigate multimode exciton–photon coupling in visible semiconductor waveguides supporting several transverse electric modes. Using rigorous coupled-wave analysis combined with a coupled-oscillator model, we show that the photonic band structure can be engineered across a range of regimes, from conventional multimode strong coupling to the superstrong coupling regime, where the Rabi coupling becomes comparable to the spacing between adjacent photonic modes. In the latter regime, hybridization between orthogonal electromagnetic modes is enabled by restricting the active material to a subregion of the mode volume where the photonic modes exhibit strong mutual overlap. This breaking of the orthogonality leads to polaritonic branches whose composition can be tuned among several photonic modes and the exciton. We demonstrate that small shifts of the exciton resonance produce pronounced changes in the propagation constants of different polariton branches, enabling exciton-controlled phase modulation through modal interference and, in the superstrong coupling regime, direct modal switching across a continuous S-shaped dispersion. The resulting figures of merit predict $\pi$ phase shifts for exciton energy shifts of only a few meV over propagation lengths of tens of micrometers, while larger shifts are still required for mode switching. These results establish multimode waveguide polaritons as a versatile platform spanning multiple coupling regimes, for compact phase and intensity control in integrated photonic architectures.
\end{abstract}

\section*{Keywords}

polaritonics, superstrong light-matter coupling, planar waveguides, integrated optics, ultrafast optical switching

\newpage

\begin{figure*}[ht] 
	\centering
	\includegraphics[width=3in]{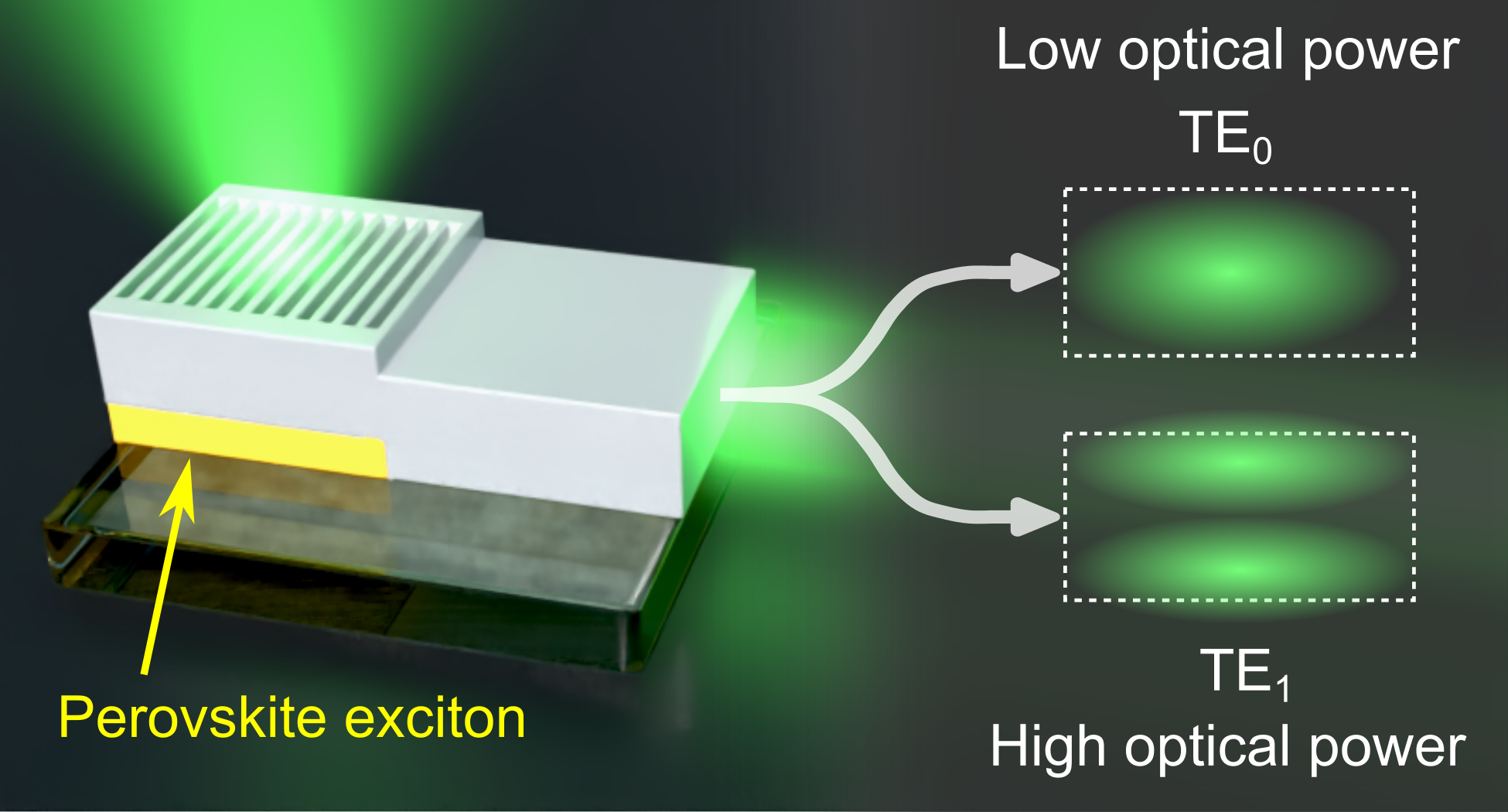}
	\caption*{TOC: Rendering of a grating-coupled planar waveguide partially filled with an excitonic material (yellow) enabling ultrafast control of the waveguide mode via the exciton nonlinearity. The resulting superstrong coupling regime enables continuous tuning between the $\mathrm{TE}_0$ and $\mathrm{TE}_1$ waveguide modes via nonlinear changes of the exciton energy depending on the optical input power.}
	\label{TOC}
\end{figure*}

\section{Introduction}

Strong coupling of light and matter has become a ubiquitous phenomenon over the past decades, driven by advances in fabrication, new material discoveries, and innovative photonic designs across a broad range of physical systems. This development led to ever higher coupling strengths between photonic modes from the microwave to visible spectral range and a multitude of coupling regimes and matter excitations, including excitons, phonons, atoms, intersubband transitions, plasmons, and magnons~\cite{Dovzhenko2018,Hertzog2019,Ballarini2019}.
Such a large light-matter interaction provides experimental access to several coupling regimes with unprecedented physical effects~\cite{FriskKockum2019,Forn-Diaz2019}, such as modifications of the vacuum ground state~\cite{DeLiberato2007,Ashhab2010,DeLiberato2017}, light-matter decoupling~\cite{DeLiberato2014,Mueller2020}, or altered chemical and material properties~\cite{Thomas2019,Garcia-Vidal2021,Cortese2021,Appugliese2022,Baydin2025}. Of particular interest for integrated photonics is the case in which a single material resonance is strongly coupled to multiple photonic modes, producing a multimode polariton dispersion whose branches inherit distinct field profiles and propagation properties from their photonic parents while sharing a common excitonic component. The resulting dispersion provides a rich design space: distinct polariton branches can be co-excited at the same energy, individually addressed by their wavevector, and shifted differentially by perturbations of the matter resonance.
Engineering such multimode dispersions across different coupling regimes is the focus of the present work. These regimes range from conventional multimode strong coupling, where each polariton branch retains a dominantly single-mode photonic character, to the so-called superstrong coupling (SSC) regime. In the latter, the Rabi couplings of two or more photonic modes coupled to the same active material become comparable to their mutual energy spacing, leading individual polaritonic branches to acquire contributions from multiple photonic modes.

The SSC region of this design space deserves particular attention because, by enabling coherent hybridization of multiple photonic modes within a single polariton branch~\cite{Mornhinweg2024,Paschos2017}, it allows for a dynamical, sub-wavelength, real-space modulation of the field intensity~\cite{Cortese2023}. Importantly, SSC can be reached by designing closely spaced photonic modes, without requiring coupling strengths approaching the bare material transition energy, as in the ultrastrong coupling (USC) regime~\cite{Ciuti2005,Anappara2009}, or the exciton binding energy, as in the very strong coupling (VSC) regime~\cite{Khurgin2001,Brodbeck2017}.

SSC was first theoretically described in 2006 for cold atoms trapped in a cavity resonator~\cite{Meiser2006}, followed by an experimental realization using nearly-degenerate transverse modes of a cavity coupled to an atomic ensemble in 2013~\cite{Wickenbrock2013}. Subsequently, more sophisticated realizations emerged in circuit and cavity QED systems in the microwave domain, coupling longitudinal cavity modes to transmon qubits~\cite{Kuzmin2019a, Jouanny2025, Sundaresan2015}, acoustic resonators~\cite{Han2016}, or magnon resonances~\cite{Kostylev2016}.

All of these realizations have in common the use of long cavities to achieve a small free spectral range, thus reducing the need for very high coupling strengths but requiring the use of high-order modes (mode orders 50-1000) to match the frequency of the matter excitation. However, neighboring high-order modes feature very similar optical properties and field profiles, which makes tuning between such states less interesting from an application perspective. At the other end of the spectrum, realizations in the THz domain use cyclotron resonances of the two-dimensional electron gas in GaAs quantum wells, which provide ultrahigh coupling strengths and thereby enable SSC with low-order modes (mode orders 1-2)~\cite{Tay2025, Mornhinweg2024a}. However, the requirement for cryogenic temperatures and high magnetic fields limits the broader applicability of these systems.

The visible domain currently presents an interesting gap in the landscape of multimode polariton realizations, as it is the domain of many promising room-temperature excitonic materials for polaritonics such as perovskites~\cite{Fieramosca2019}, TMDs~\cite{Hu2020}, and organic dyes~\cite{Jiang2022}. To date, SSC realizations in this spectral range have focused exclusively on microcavities~\cite{Adl2025, Mandal2023, Georgiou2021, Godsi2023}. While these works achieved SSC with medium-order modes (mode orders~5-12), coupling of the two lowest-order cavity modes has thus far remained elusive due to the unfavorable energy spacing between them: an exciton resonant with the fundamental cavity mode at a given energy would only encounter the second mode at twice that energy because cavity energies depend approximately linearly on the mode order. Reaching SSC with the lowest-order cavity mode would therefore demand coupling strengths in the USC regime, which are difficult to achieve in the visible range. 

Here, we address this limitation by replacing microcavities with planar waveguide modes, accessed by a 1D grating, to enable multimode polariton dispersion engineering between the $\mathrm{TE}_0$ and $\mathrm{TE}_1$ modes and a room-temperature exciton resonance in the visible spectral range. In contrast to cavities, the mode energies in waveguides are governed by effective refractive indices, which depend on the waveguide thickness and the extent of the evanescent field but only weakly on mode order. In particular, all effective indices converge toward the bulk refractive index for sufficiently thick waveguides, enabling arbitrarily small energy spacings between low-order modes. The waveguide thickness thus serves as a single design knob that tunes the system continuously from multimode strong coupling to the SSC regime, hybridizing the distinct field profiles of the two lowest-order modes. 

\begin{figure*}[b!] 
	\includegraphics[width=170mm]{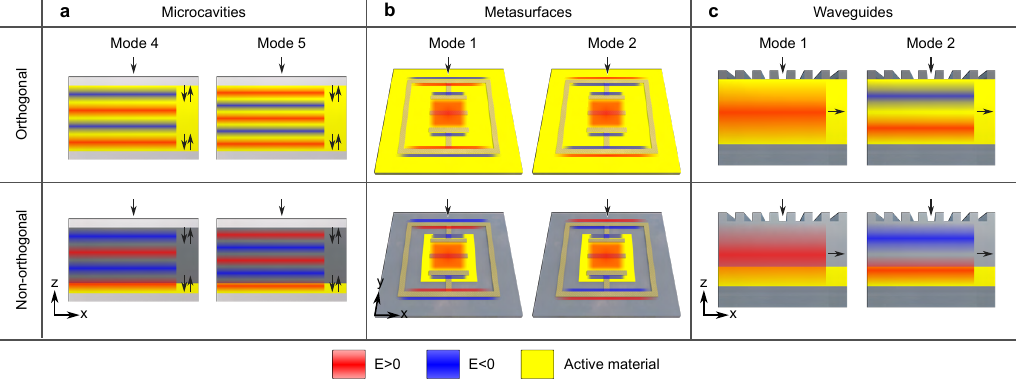}
	\caption{Comparison of platforms for superstrong polariton physics. When the active material (yellow) fills the entire mode volume (upper row), photonic modes are orthogonal over the active region in a Hermitian system, resulting in polariton modes containing only a single photonic mode. Localizing the active material within part of the waveguide (lower row) induces a nonzero overlap between photonic modes over this region, enabling polariton modes with contributions from multiple photonic modes and allowing continuous, coherent tuning between them. (a)~Microcavities typically require coupling between higher-order modes to reach the SSC regime, due to the large energy spacing between low-order modes. (b)~Metasurfaces can be engineered for high overlap, but fabrication can be more complex. (c)~Waveguides provide both high overlap and low energy separation between low-order modes, making them ideal for SSC. Blue and red overlays indicate the sign of the electric field, illustrating that positive and negative contributions to the overlap integral cancel in the case of full filling, while partial filling allows for a nonzero overlap.}
	\label{Fig1}
\end{figure*}

For the theoretical description of the waveguide structures, we use a generalized Hopfield model introduced in Ref.~\cite{Cortese2023}, which has been successfully applied to other photonic systems~\cite{Mornhinweg2024a,Tay2025}. This work showed how the conditions required to reach SSC can be geometrically engineered. Different electromagnetic modes are in fact orthogonal and therefore induce orthogonal exciton polarization distributions when the full mode volume is occupied by the active material (top row in~\cref{Fig1}). In this case, the exciton polarization distributions remain independent of each other, so that each polariton contains only a single photonic component, making SSC impossible.
Creating polaritons with contributions from multiple photonic modes instead requires a high spatial overlap of the photonic mode profiles over the region occupied by the active material. This can be realized by restricting the active material to a subregion of the mode-volume (bottom row of~\cref{Fig1}), thereby breaking the orthogonality of the modes over the active region and enabling coupling through the non-orthogonal polarization distributions of the excitons~\cite{Cortese2023}. Early works on SSC naturally achieved these high overlaps since the matter excitation was confined to a negligibly small portion of the long cavities, and therefore this aspect was not discussed explicitly (see Table~S1 in the Supporting Information for an overview of representative SSC systems). 
In contrast, the visible-range implementations we target here require larger fractions of the mode volume to be filled with the active material, making a more careful placement essential, as discussed later. The resulting SSC regime is distinguished, among other features, by a characteristic S-shaped dispersion of the hybrid polariton branch connecting two distinct photonic modes (discussed later in the context of \cref{Fig2,Fig3_2}). We note that while high modal overlap is a geometric property accessible at any coupling strength, its physical consequences, coherent hybridization of multiple photonic modes within a single polariton branch, only become manifest when the coupling strength reaches the SSC regime. Below SSC, the same geometry instead produces a multimode strong coupling dispersion with degenerate branches that can be co-excited by a broad-wavevector source, as we exploit in the use cases discussed below. 

We note that apart from the waveguides discussed here, SSC with low-order modes can also be realized using metasurfaces~\cite{Mornhinweg2024a}, as summarized in~\cref{Tab_ComparisonSystems}. However, achieving a high modal overlap in metasurfaces typically requires in-plane patterning of the active material by etching, a process that risks damaging the delicate excitonic materials and adds significant fabrication complexity. Planar waveguides, by contrast, employ the same simple layered deposition used for cavities: the active material is deposited as a thin layer followed by a dielectric spacer, without any etching steps on the active layer itself. An overview of these three platforms is provided in~\cref{Fig1} and~\cref{Tab_ComparisonSystems}.

\begin{table}[t!]
	\centering
	\footnotesize
	\begin{tabular}{@{} l c c c @{}}
		\toprule		
		\textbf{Feature} & \textbf{Microcavities} & \textbf{Metasurfaces} & \textbf{Waveguides} \\
		\midrule
		Modeling              & \cmark{} Simple (1D)         & \xmark{} Advanced 3D vectorial           & \cmark{} Simple (1D/2D) \\
		Fabrication & \cmark{} Layered, no etching & \xmark{} Etching, risk of damage & \cmark{} Layered, no etching \\
		Mode energy separation           & \xmark{} Fixed (1 octave)    & \cmark{} Tunable                & \cmark{} Tunable \\
		Q-factor control                 & \cmark{} via mirror reflectivity             & \cmark{} separate for each mode             & \cmark{} via grating thickness \\
		Coupling of low-order modes   & \xmark{} generally not possible    & \cmark{} Possible               & \cmark{} Possible \\
		\bottomrule
	\end{tabular}
	\caption{Comparison of platforms for superstrong polariton physics.}
	\label{Tab_ComparisonSystems}
\end{table}

In this work, we investigate planar waveguides across a range of multimode coupling regimes, using rigorous coupled-wave analysis (RCWA) with realistic parameters of commonly used two-dimensional hybrid organic-inorganic perovskites. By varying the waveguide thickness we tune the photonic mode spacing and thereby traverse the regime from multimode strong coupling to SSC of the two lowest-order modes. By fitting the generalized Hopfield model from Ref.~\cite{Cortese2023} to the simulated dispersions, we extract the overlap coefficients and Hopfield coefficients of the polariton modes, confirming the characteristic S-shaped dispersion and the presence of multiple photonic components within individual polariton branches. Building on these results, we demonstrate two complementary use cases~\cite{Ezratty2025}: an interference-based phase modulator that operates in the multimode strong coupling regime, and an exciton-tunable modal switch that exploits the avoided crossing characteristic of the SSC regime.

\section{Results}

\subsection{Theoretical Framework}

To describe the coupling of multiple photonic modes to a single exciton in our waveguides, we use the recently reported Hopfield-type model introduced in Ref.~\cite{Cortese2023}, briefly summarized here for the case of $N$ photonic modes with coupling strengths below the USC regime. A key feature of this model is the introduction of an overlap parameter $\eta_{jk}$ quantifying the normalized spatial overlap between the electric fields $\vb{F}_j$ and $\vb{F}_k$ of photonic modes $j$ and $k$ over the active material region $\mathcal{V}_\mathrm{act}$ as:

\begin{equation} \label{Eq_Overlap}
	\eta_{jk} =
	\frac{
		\displaystyle
		\int_{\mathcal{V}_\mathrm{act}} \mathrm{d}^3\mathbf{r}\;
		\vb{F}_{j}^{*}(\mathbf{r}) \cdot \vb{F}_{k}(\mathbf{r})
	}{
		\sqrt{
			\left(
				\displaystyle
				\int_{\mathcal{V}_\mathrm{act}} \mathrm{d}^3\mathbf{r}\;
				|\vb{F}_{j}(\mathbf{r})|^2
			\right)
			\left(
				\displaystyle
				\int_{\mathcal{V}_\mathrm{act}} \mathrm{d}^3\mathbf{r}\;
				|\vb{F}_{k}(\mathbf{r})|^2
			\right)
		}
	}.
\end{equation}
An accurate description of arbitrary mode overlap $\eta_{jk}$ requires accounting for the coupling between photonic modes mediated by the exciton. To achieve this, the model introduces $N$ degenerate exciton states as auxiliary degrees of freedom, serving as a basis for capturing the effect of the overlap on the coupling between the photonic modes and the single physical exciton state. The resulting Hamiltonian therefore contains $2N$ states ($N$ photonic and $N$ excitonic), which can be expressed as:

\begin{align} \label{Hamiltonian}
H =
\begin{pmatrix}
	\mathbf{E}_{\mathrm{ph}} & \mathbf{G} \\
	\mathbf{G}^\mathrm{\dagger} & \mathbf{E}_{\mathrm{ex}}
\end{pmatrix},
\end{align}

where the submatrices are given by:

\begin{equation}
\begin{array}{l}
	\mathbf{E}_{\mathrm{ph}} = \mathrm{diag}(E_{\mathrm{ph},1},\ldots,E_{\mathrm{ph},N}), \\[0.5em]
	\mathbf{E}_{\mathrm{ex}} = \mathrm{diag}(E_{\mathrm{ex}},\ldots,E_{\mathrm{ex}}),
\end{array}
\quad
\mathbf{G} =
	\begin{pmatrix}
		g_{11} & 0      & \cdots & 0 \\
		g_{21} & g_{22} & \cdots & 0 \\
		\vdots & \vdots & \ddots & \vdots \\
		g_{N1} & g_{N2} & \cdots & g_{NN}
	\end{pmatrix},
\end{equation}
where $E_{\mathrm{ph},j}$ are the energies of the photonic modes, $E_{\mathrm{ex}}$ is the exciton energy, and $g_{jk}$ are the coupling strengths between photonic mode $j$ and degenerate exciton state $k$. Note that the lower-triangular shape of the coupling strength matrix $\mathbf{G}$ is a result of the choice of a specific basis for the exciton states, but other forms would be equally valid. The physical exciton fraction of a polariton mode is determined by the sum over all exciton states and is therefore independent of the choice of basis.

The coupling strength matrix $g_{jk}$ can be related to the previously calculated overlap coefficients via an expansion coefficient matrix $\alpha_{jk}$, which determines how the total coupling strength of each photonic mode $j$ to the physical exciton $g_j$ is distributed over the degenerate exciton states:

\begin{align} \label{Hamiltonian_DefiningEq}
	g_{jk} = g_j \, \alpha_{jk} \quad \text{with} \quad \alpha_{jk} = 0\ \text{for}\ k > j,\ \text{and}\ \eta_{jk} = \sum_{\ell=1}^{\min(j,k)} \alpha_{j\ell}^* \, \alpha_{k\ell}.
\end{align}

This definition automatically ensures normalization of the expansion coefficients as $\sum_{k=1}^{j} |\alpha_{jk}|^2 = 1$, and therefore $|\alpha_{jk}| \in [0,1]$. The explicit $N=2$ form of the model is provided in the Methods section. 

Overall, this model generalizes the two commonly used limiting cases, the so-called $2N$ and $N\!+\!1$ models~\cite{Adl2025,Godsi2023, Balasubrahmaniyam2021, Georgiou2021}. When the active material fills the entire mode volume, the overlap matrix reduces to $\eta_{jk} = \delta_{jk}$, recovering the $2N$ model, where each photonic mode couples to an independent degenerate exciton state. Instead, for an ideal scenario with unit overlap $\eta_{jk} = 1$, it recovers the $N\!+\!1$ model, where all photonic modes couple to a single exciton, leaving $N-1$ dark, uncoupled modes.

The model can therefore describe various multimode coupling regimes, with SSC being characterized by at least one pair of photonic modes $(j,k)$ whose coupling to the same exciton state $l$ is comparable to or larger than their energy spacing $\Delta E_{\mathrm{ph},jk}$:
\begin{align} \label{EqSSC_Condition}
g_{jl}, g_{kl}	\simeq \Delta E_{\mathrm{ph},jk} = |E_{\mathrm{ph},k} - E_{\mathrm{ph},j}|
\end{align}
We stress here that as SSC describes the breaking of the single-mode polaritonic approximation, its violation is a smooth process. As such, the above condition serves as a useful reference for assessing whether the SSC regime can be realized in a specific device, while definitive evidence of SSC is provided by non-negligible contributions from two or more different photonic modes to the same polariton eigenstate, as analyzed in \cref{Fig2,Fig3_2}.
	
One way to fulfil the SSC condition, is through a high spatial overlap ($\eta_{jk} \approx 1$) in combination with high total coupling strengths $g_j, g_k \simeq \Delta E_{\mathrm{ph},jk}$. Such a high overlap directly shapes the polariton dispersion. In the perfect overlap limit of ${\eta_{jk} = 1}$, the energy gap between pairs of upper and lower polariton branches originating from different photonic modes closes, yielding a characteristic S-shaped dispersion in which it is possible to continuously tune the polariton from one dominant photonic character to the other (see an example of such a dispersion in~\cref{Fig2}). In the realistic cases considered here, where $\eta_{jk} < 1$, a finite gap $\Delta E_{\mathrm{pol},jk}$ opens at the crossover between polaritons $j$ and $k$, but remains insignificant as long as it is small compared with the linewidths of the involved polaritons, in our case dominated by the exciton linewidth $\gamma_{\mathrm{ex}}$. As shown for the $N = 2$ case in Section~S5 of the Supporting Information, 
a sufficiently small energy gap can be realized both by large overlaps or by low coupling strengths. Therefore, it is not possible to define a general overlap threshold for achieving continuous tuning in the SSC regime.

In the following, we fit the model to the polariton dispersion from RCWA simulations with $E_{\mathrm{ex}}$ and $E_{\mathrm{ph},j}$ fixed, using a total of $N(N+1)/2$ fitting parameters: $N$ real coupling strengths $g_j$ and $N(N-1)/2$ real-valued $\alpha_{jk}$ parameters with the normalization ensured by a hyperspherical parameterization. Note that only bound waveguide modes are included, as the quantities extracted from the model depend only on the fields within the waveguide core, where the guided-mode basis provides an accurate description. A complete reconstruction of the electromagnetic field outside the core would additionally require including the radiation continuum. The eigenvectors yield the Hopfield coefficients (fractions) for the photonic modes $|C_j|^2$ and the total exciton fraction $|X|^2 = \sum_{j=1}^N |X_j|^2$ of each polariton, allowing us to confirm the presence of multiple photonic modes inside a polariton mode as well as the possibility to continuously tune between them.

We now apply this general theoretical framework to the case of a planar waveguide geometry.

\subsection{Ideal Waveguide Model}

To gain an understanding of how to achieve high modal overlap, we first analyze an idealized system amenable to an analytical treatment: two waveguide modes in a planar waveguide of total thickness $D$ with perfectly reflecting walls (see Section~S6 of the Supporting Information 
for details), resulting in the following field profiles:

\begin{equation} \label{Eq_IdealWaveguideModel}
	\vb{F}_j(x, z) = \sin\left(\frac{j \pi z}{D}\right) \, \hat{\vb{y}}.
\end{equation}

Note that we use $j=1$ for the $\mathrm{TE}_0$ mode and $j=2$ for the $\mathrm{TE}_1$ mode throughout, i.e., the index $j$ corresponds to the $\mathrm{TE}_{j-1}$ waveguide mode.

We optimize both the position and the volume fraction $\Delta z / D$ of the active material layer of thickness $\Delta z$ (see geometry in \cref{Fig2}(a)). The optimization is performed by directly calculating the overlap integral of the electric fields $\vb{F}$~(\cref{Eq_Overlap}) to maximize the geometric mean of the coupling strengths, $\sqrt{g_1 g_2}$, for a given target overlap $\eta_{12}$.

The results of this optimization reveal a simple design principle: for maximizing the coupling strength at any desired overlap, the optimal placement for a single layer of active material is either at the bottom or at the top of the waveguide. The thickness of this layer then determines the overlap, with thinner layers yielding higher overlaps. Intuitively, for an infinitesimally thin layer at the waveguide boundary, the overlap approaches unity because ideal waveguide modes vanish at the boundaries and can be approximated as linear functions with different slopes. In this limit, the modes are identical up to a scaling factor, resulting in unity overlap, which we remind is a normalized, dimensionless quantity.

However, as the field strength is low near the outermost positions, the coupling strength is also low, thus requiring a trade-off between overlap and coupling strength. An active material fraction of $\Delta z / D = 0.3$ was chosen for the numerical simulations, providing an overlap of $\eta_{12} = 0.99$ while still maintaining a high coupling strength.

\subsection{Waveguide Design and Simulation}

Based on these theoretical considerations, we design a grating-coupled planar waveguide structure to achieve SSC in the visible range. The key advantage of waveguides is that the photonic mode energy spacing can be tuned without significantly affecting the energy of the fundamental $\mathrm{TE}_0$ mode. Specifically, in the limit where the waveguide is much thicker than the wavelength (see Methods section
), the mode energies can be approximated as:
\begin{equation} \label{EqWaveguideSpacing}
E_j(\beta) \approx \frac{\hbar c}{n_0} \beta \left[1 + \underbrace{\frac{\pi^2 n_0^2}{2 \beta^2} \frac{j^2}{D^2}}_{\ll 1}\right],
\end{equation}
where $n_0$ is the refractive index of the waveguide material, and $\beta$ is the in-plane wavevector (propagation constant) of the waveguide mode. Note that the mode energy depends only weakly on the mode order $j$, while the energy spacing between modes scales as $1/D^2$ and can therefore be tuned as needed, down to the point of near-degeneracy.
In contrast to microcavities where the mode spacing is fixed to the fundamental mode energy, this enables exploring SSC with the two lowest-order modes ($\mathrm{TE}_0$, $\mathrm{TE}_1$), which offers two advantages: (1)~the modes have more distinct field profiles and thus more distinct optical properties; and (2)~as shown later in~\cref{Fig4}, for a given desired overlap, the achievable coupling strength is significantly higher than for higher-order mode pairs.

\begin{figure*}[ht]
	\includegraphics[]{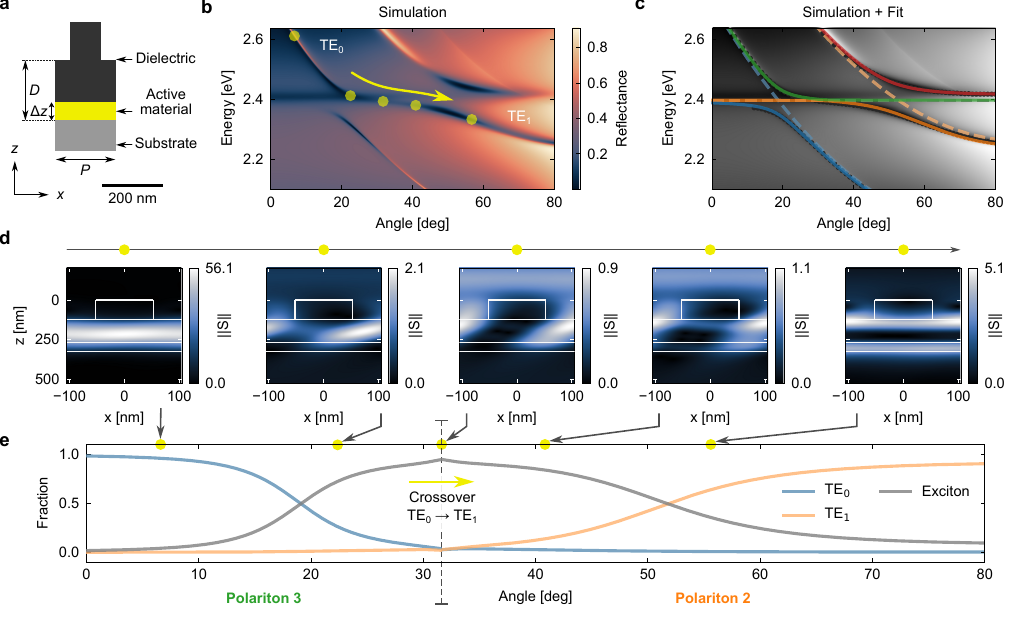}
	\caption{Simulation of multimode strong coupling with high modal overlap in the low-thickness waveguide ($D=200$~nm). (a)~Unit cell of the grating structure with active material fraction $\Delta z/D = 0.3$ and periodicity $P=210$~nm. (b)~Polariton dispersion in reflection. (c)~Fit using $\mathrm{TE}_0$ (blue dashed) and $\mathrm{TE}_1$ (orange dashed) photonic modes, yielding four polariton branches. Small gray dots below the colored curves indicate points used in fit. (d)~Poynting vector magnitude at selected yellow points in (b), illustrating the crossover from $\mathrm{TE}_0$ to $\mathrm{TE}_1$. (e)~Fractions of uncoupled modes. Polariton~2 (orange) is shown to the right and polariton~3 (green) to the left of the crossover line (gray dashed), which marks the angle where the energy gap between the polaritons is minimal. Due to the small energy gap, no significant discontinuity is observed in the fractions. All evaluated quantities of this waveguide are listed in \cref{Tab_WaveguideComparison}.}
	\label{Fig2}
\end{figure*}

In order to demonstrate the tunability of the mode energy spacing, we focus on three different waveguide geometries with different core thicknesses in the rest of the paper: low- ($D=200$~nm), mid- ($D=400$~nm), and high-thickness ($D=600$~nm) waveguides. Each thickness serves a distinct role in the analysis that follows: the 200~nm waveguide illustrates the multimode strong coupling regime below SSC and underpins the interference-based phase modulator
; the 400~nm waveguide reaches the SSC regime proper and underpins the modal-switching scheme
; and the 600~nm waveguide supports five photonic modes simultaneously and is used for the systematic parameter sweep over active-material fraction and mode order in~\cref{Fig4}. As shown in~\cref{Fig2}(a), we place the active material of thickness $\Delta z$ as bottom-most layer of the waveguide core, comprising a fraction $\Delta z / D = 0.3$ of the total core thickness, as motivated by the optimization results for ideal waveguides. As parameters for our active material, we choose values similar to the A-exciton in common two-dimensional hybrid organic-inorganic perovskite crystals~\cite{Song2021} with a resonance in the visible at 517 nm ($E_\mathrm{ex} = 2.398$~eV). We note that with these parameters the $\mathrm{TE}_0$ and $\mathrm{TE}_1$ modes of all studied waveguides fulfill the strong coupling condition, as verified in \cref{Tab_WaveguideComparison}. For coupling light into the waveguide modes, we consider a one-dimensional grating on top of the waveguide core. More details on the materials and geometric parameters are available in the Methods section.

We then use rigorous coupled-wave analysis (RCWA, S4~\cite{Liu2012}) to simulate the reflectance of the structure as a function of incidence angle and extract the polariton dispersion. Details of the simulation and extraction procedure are provided in Section~S3 of the Supporting Information. 
The resulting dispersion is fitted using the model described in the previous section to retrieve the overlap coefficients and Hopfield coefficients of the polariton modes.

In this context, we also point out that the number of photonic modes required for an accurate theoretical description is naturally limited for two reasons. First, the overlap integral between the fundamental mode and a higher-order mode $j$ remains large only for modes whose field varies slowly across the active layer, i.e., for $j \lesssim D / \Delta z$. Second, according to \cref{EqWaveguideSpacing}, the energy spacing between adjacent waveguide modes increases approximately linearly with $j$, in contrast to the constant energy spacing of microcavities. As a result, the contribution of progressively higher-order modes is suppressed both by their decreasing overlap with the active layer and by their increasing energetic detuning.

\subsection{Superstrong Coupling in Simulated Structures}

We start by analyzing the thin waveguide ($D = 200$~nm), deliberately chosen to yield a large mode energy spacing that places the system in the multimode strong coupling regime below SSC, with $g_{11}/\Delta E_{\mathrm{ph},12} = 0.11$ and $g_{21}/\Delta E_{\mathrm{ph},12} = 0.12$. The simulation results in~\cref{Fig2} exhibit the characteristic S-shaped dispersion, enabling continuous tuning between polariton modes dominated by $\mathrm{TE}_0$ and $\mathrm{TE}_1$ as the incidence angle is varied.

Fitting the Hopfield model ($N = 2$) produces four polariton branches that closely match the simulated dispersion, with a root-mean-square error of $3.6$~meV per point, well below the linewidths of the relevant modes (see~\cref{Tab_WaveguideComparison}). The high overlap $\eta_{12} = 0.99$ results in a negligibly small energy gap of $3$~meV between the two inner polariton branches (labeled 2 (orange) and 3 (green) in~\cref{Fig2}), well below the polariton linewidths. This high overlap, combined with the moderate coupling strength, places the system in a multimode strong coupling regime: the dispersion exhibits avoided crossings, but the individual polariton branches remain dominantly composed of a single photonic mode, wherever the photonic fraction is nonnegligible, as confirmed by the Hopfield coefficients in~\cref{Fig2}(e).

\begin{figure*}[ht]
	\includegraphics[]{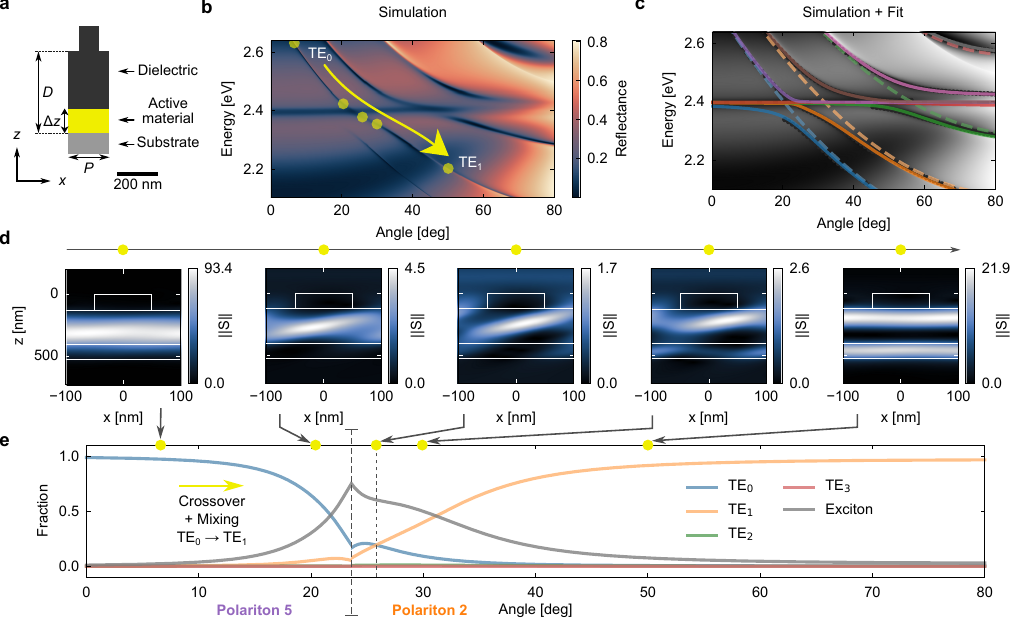}
	\caption{Simulation of superstrong coupling in the mid-thickness waveguide ($D=400$~nm). (a)~Unit cell of the grating structure with active material fraction $\Delta z/D = 0.3$ and periodicity $P=200$~nm. (b)~Polariton dispersion in reflection. (c)~Fit using $\mathrm{TE}_0$-$\mathrm{TE}_3$ photonic modes (dashed lines), yielding 8 polariton branches. Small gray dots below the colored curves indicate points used in fit. (d)~Poynting vector magnitude at selected yellow points in (b), illustrating the crossover from $\mathrm{TE}_0$ to $\mathrm{TE}_1$. (e)~Fractions of uncoupled modes. Polariton~2 (orange) is shown to the right and polariton~5 (purple) to the left of the crossover line (gray dashed), which marks the angle where the energy gap between the polaritons is minimal. Due to the small energy gap, no significant discontinuity is observed in the fractions. All evaluated quantities of this waveguide are listed in \cref{Tab_WaveguideComparison}.}    
	\label{Fig3_2}
\end{figure*}

As a consequence of their non-SSC nature, the polariton modes are dominantly exciton-like at the crossover point, leading to weak electromagnetic field confinement (see maximal Poynting vector values in~\cref{Fig2}(d)). This device will be exploited for the interference-based phase modulation
, where the two branches are co-excited by a broad-wavevector source.
Increasing the waveguide thickness while keeping the active-material fraction fixed at 30\% reduces the mode energy spacing without significantly affecting the overlap. For the mid-thickness waveguide ($D = 400$~nm,~\cref{Fig3_2}), this increase in thickness therefore drives the system into the SSC regime with $g_{11}/\Delta E_{\mathrm{ph},12} = 0.26$ and $g_{21}/\Delta E_{\mathrm{ph},12} = 0.37$, while maintaining the high overlap of $\eta_{12} = 0.97$. Due to the higher coupling strength, the energy gap between the two inner polariton branches at the $\mathrm{TE}_0$-$\mathrm{TE}_1$ crossover point increases to $13.8$~meV, which is still below the polariton linewidths, dominated by the exciton linewidth $\gamma_{\mathrm{ex}} = 30$ meV, and thus sufficiently small for allowing a continuous transition.

Analysis of the Hopfield coefficients of the polariton modes confirms the presence of both photonic modes in each polariton branch, as well as the continuous tuning between them across the crossover point (see~\cref{Fig3_2}(e)). Furthermore, the increased coupling strength results in a significantly higher photonic fraction at the crossover point, leading to better field confinement and higher reflection contrast. Contrary to the thin waveguide, this thicker geometry therefore allows the creation of coherent superposition states of the two photonic modes in reflection, even with an incoherent excitation source, with the coherence provided by the mutual coupling to the exciton.

We note that in thicker waveguides, the $\mathrm{TE}_0$-dominated polariton mode profile slightly retreats from the active material layer at energies above the exciton resonance (see~\cref{Fig3_2}(d), second field profile). This effect arises because the real part of the refractive index of the active material becomes lower than that of the dielectric filler near the exciton resonance, causing the active layer to act as a low-index region and thus repelling the mode. Importantly, this shift does not significantly affect the coupling strength, which remains comparable to that of the thin waveguide (cf.~\cref{Tab_WaveguideComparison}).

\begin{figure*}[ht]
	\includegraphics[]{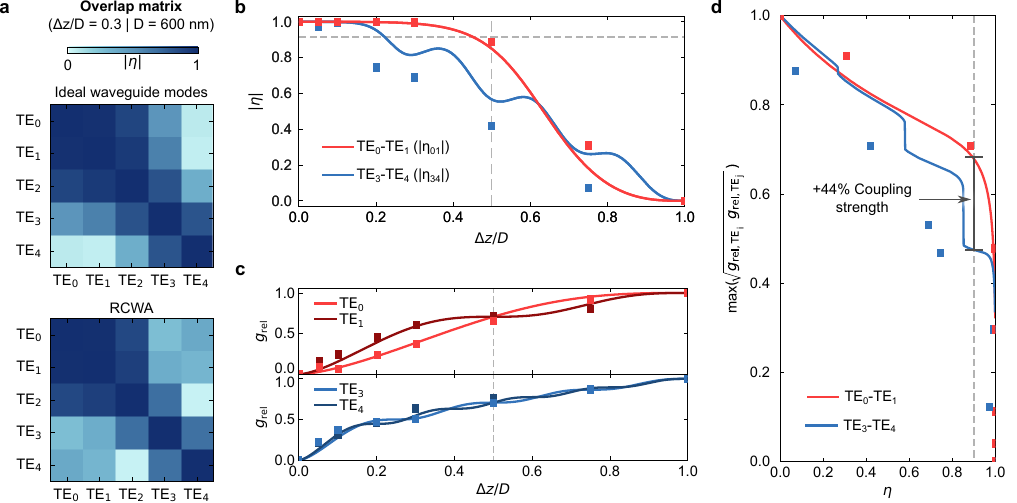}
	\caption{Impact of active material fraction on crossover behavior for the high-thickness waveguide ($600$~nm). (a)~Example of an overlap matrix $\eta_{jk}$ at an active material fraction $\Delta z / D = 0.3$. All evaluated quantities of this waveguide are listed in \cref{Tab_WaveguideComparison}. (b)~Overlap between $\mathrm{TE}_0/\mathrm{TE}_1$ and $\mathrm{TE}_3/\mathrm{TE}_4$ decreases with increasing active material fraction $\Delta z / D$. (c)~Conversely, the total relative coupling strengths $g_{\mathrm{rel},j}(\Delta z) = g_j(\Delta z)/g_j(D)$ of photonic modes to excitons increase. (d)~Maximally achievable geometric mean of relative coupling strengths for a given desired overlap for ideal waveguide modes (\cref{Eq_IdealWaveguideModel}), highlighting an increased coupling strength for $\mathrm{TE}_0/\mathrm{TE}_1$ modes compared to higher-order modes. All values shown in this figure are obtained using the Hamiltonian model, with lines indicating results using \cref{Eq_IdealWaveguideModel}, and rectangles indicating fits to simulated RCWA data. Full fit results and additional details provided in Section~S6 of the Supporting Information.}
	\label{Fig4}
\end{figure*}

Lastly, we investigate how the fraction of active material influences the crossover behavior and compare the performance of coupling the lowest-order modes ($\mathrm{TE}_0$/$\mathrm{TE}_1$) with that of higher-order modes ($\mathrm{TE}_3$/$\mathrm{TE}_4$). For this analysis, we use the thick ($D=600$~nm) waveguide, which enables simultaneous observation of the dispersion for $\mathrm{TE}_0$ - $\mathrm{TE}_4$ modes. The results of the fits for various active material fractions $\Delta z / D$ ranging from $0.05$ to $1$ are presented in~\cref{Fig4}, with the complete characterization equivalent of \cref{Fig2,Fig3_2} provided in Section~S4.3 of the Supporting Information.

The theoretical model of ideal waveguide modes provides a reasonable estimate of the overlap matrix $\eta_{jk}$, with the fitted overlap values following the direct calculation of the overlap integral using \cref{Eq_IdealWaveguideModel} (see~\cref{Fig4}(a)). This agreement confirms both the validity of the theoretical model and the successful extraction of the overlaps from the fit. For the active material fraction $\Delta z / D = 0.3$, we find $\eta_{12} = 0.995$ (ideal: $0.990$), $\eta_{14} = 0.879$ (ideal: $0.893$), and $\eta_{45} = 0.687$ (ideal: $0.816$), with values for other active material fractions shown in~\cref{Fig4}(b).
The ideal waveguide model~(\cref{Eq_IdealWaveguideModel}) also predicts the coupling strength, normalized to the value for a completely filled waveguide, $g_\mathrm{rel}$, as a function of the active material fraction (see~\cref{Fig4}(c)), using that $g_j \propto \sqrt{\int \mathrm{d}z \, |\vb{F}_j(z)|^2}$. As shown in~\cref{Fig4}(c), these predictions closely match the values extracted from the RCWA fits. In this context, we also note that restricting the Hamiltonian to the first five waveguide modes is justified, even for the high ratio of coupling strength to mode spacing $g_2/ \Delta E_{\mathrm{ph},12} = 0.74$ in the thick waveguide. As demonstrated in Figures S4 and S5 of the Supporting Information, each of the ten polariton branches contains, to good approximation, contributions from no more than two photonic modes at any given detuning. Therefore, modes of even higher order can be safely neglected.

Two important results emerge from this analysis. (1)~There is a fundamental tradeoff between coupling strength and overlap: increasing the active material fraction increases the coupling strength but reduces the overlap. (2)~For a given desired overlap, the achievable coupling strength can be significantly higher when using the lowest-order modes ($\mathrm{TE}_0$/$\mathrm{TE}_1$) compared to higher-order modes ($\mathrm{TE}_3$/$\mathrm{TE}_4$), as shown in~\cref{Fig4}(d). In particular, to achieve a high overlap of $\eta_{jk} = 0.9$, the maximally achievable coupling strength for $\mathrm{TE}_0$/$\mathrm{TE}_1$ modes is $44\%$ higher than for $\mathrm{TE}_3$/$\mathrm{TE}_4$ modes under ideal placement of the active material. This is because the rapid spatial oscillations of higher-order modes lead to cancellation of positive and negative contributions to the overlap integral, thus requiring a thinner layer to achieve the same overlap, which in turn reduces the coupling strength. Note that these two results apply equally to cavities, as their idealized spatial mode profiles are identical to those of waveguides. However, coupling between $\mathrm{TE}_0$ and $\mathrm{TE}_1$ modes is generally not feasible in cavities due to their large mode energy spacing, which highlights a key advantage of waveguides in realizing superstrong coupling.

\section{Discussion}
\label{Sec_Applications}

 In this section we explore different use cases for the multimode waveguide platform we have described and characterized.
 The first relies on interference between two polariton modes in the multimode strong coupling regime to achieve phase modulation by shifting the exciton resonance. The second exploits switching between two photonic modes by sweeping the exciton energy across the avoided crossing.

\subsection{Interference-Based Phase Modulation in the Multimode Strong Coupling Regime} \label{Sec_Applications_Interference}

We first demonstrate that even modest exciton energy shifts of less than 10~meV enable mode switching via interference between two polariton modes. The key observation is that two photonic modes strongly coupled to a single exciton result in a dispersion that allows two polariton modes to be excited at the same energy but with distinct wavevectors that respond differently to changes in $E_\mathrm{ex}$. As the polaritons propagate, they accumulate a phase difference $\Delta\Phi$ that can be tuned by adjusting $E_\mathrm{ex}$, thereby modulating their interference at the output.

\begin{figure*}[ht!]
	\centering
	\includegraphics[]{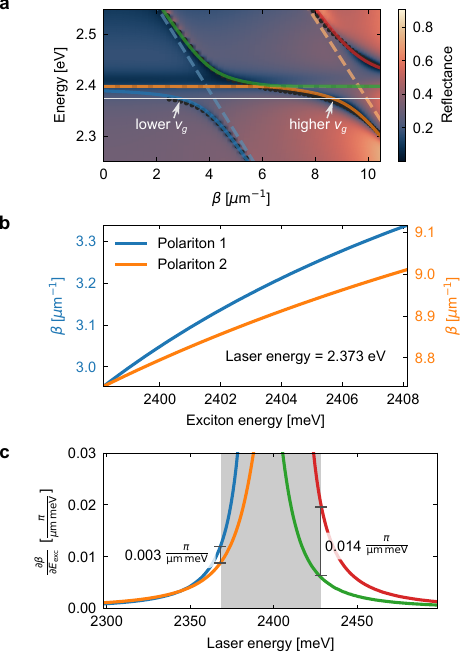}
	\caption{Exciton-tunable interference-based switching in the low-thickness waveguide ($D = 200$~nm, $\Delta z/D = 0.3$). (a)~Dispersion of the waveguide from~\cref{Fig2} as a function of in-plane wavevector~$\beta$. A focused laser at fixed energy (white horizontal line) coherently excites both polariton~1 (blue, $\mathrm{TE}_0$) and polariton~2 (orange, $\mathrm{TE}_1$) due to its broad wavevector distribution. (b)~Tuning the exciton energy $E_\mathrm{ex}$ leads to distinct shifts in $\beta$ for the two polaritons due to their different exciton fractions and group velocities. (c)~Magnitude of the wavevector shift, shown as $\partial \beta/\partial E_\mathrm{ex}$ evaluated at the original exciton energy ($2.398$~eV). The difference between these derivatives for the two polariton modes quantifies the phase shift per unit energy shift and propagation length, serving as the figure of merit (FOM) for this switching scheme. The FOM increases as the laser energy approaches the exciton resonance. The gray shaded area indicates an energy range of $2\gamma_{\mathrm{ex}}$ centered at the exciton, where the system is too lossy for operation. All evaluated quantities of this waveguide are listed in \cref{Tab_WaveguideComparison}.}
	\label{Fig6}
\end{figure*}

In~\cref{Fig6}, we illustrate this mechanism for the low-thickness waveguide. Here, a focused laser with a broad wavevector distribution would coherently excite both polariton~1 and polariton~2 at a fixed energy $E_\mathrm{Laser}$ (white horizontal line). We then evaluate how the wavevectors $\beta_1$ and $\beta_2$ of these two polariton modes shift as a function of the exciton energy $E_\mathrm{ex}$. To quantify the sensitivity of each polariton for small shifts in the exciton energy, we use the value of the slope $\partial \beta / \partial E_\mathrm{ex}$. The difference between these slopes then defines the figure of merit (FOM) for interference-based switching:
\begin{equation} \label{Eq_FOM_switching}
	\mathrm{FOM} = \left|\frac{\partial \beta_1}{\partial E_\mathrm{ex}} - \frac{\partial \beta_2}{\partial E_\mathrm{ex}}\right|, \qquad
	\Delta\Phi_\mathrm{ex} = \mathrm{FOM}\, \Delta E_\mathrm{ex}\, L,
\end{equation}
where $\Delta E_\mathrm{ex}$ is the exciton energy shift, $L$ is the fixed propagation distance in the waveguide, and $\Delta\Phi_\mathrm{ex}$ denotes the tunable phase shift caused by $\Delta E_\mathrm{ex}$, which adds to the static term $\Delta\Phi_0 = (\beta_1-\beta_2)\,L$ evaluated at $\Delta E_\mathrm{ex} = 0$.

To gain physical insight into this mechanism, we note that in the low-thickness waveguide, where the two polariton modes are predominantly composed of a single photonic mode, a $2\times 2$ Hopfield model yields the closed-form result (derived in Section~S7 of the Supporting Information):
\begin{equation} \label{Eq_2x2}
	\frac{\partial\beta_j}{\partial E_\mathrm{ex}} = -\frac{|X_j|^2}{\hbar v_{g,j}},
\end{equation}
where $|X_j|^2$ is the exciton fraction and $v_{g,j}$ is the group velocity of polariton mode $j$, shown in Figure~S17 of the Supporting Information. 
While the results from this $2\times 2$ Hamiltonian are instructive for guiding the design, we point out that we use the full multi-mode Hamiltonian~(\cref{Hamiltonian}) and directly calculate the derivatives with \cref{Eq_FOM_switching} in the following.

As shown in~\cref{Fig6}(c), the FOM increases as the laser energy approaches the exciton resonance. This enhancement is driven by both the higher excitonic fraction and the reduced group velocity of the polariton modes near resonance, as described by~\cref{Eq_2x2}. In order to avoid the excessive losses from the exciton, we determine the FOM at a laser detuning of $E_\mathrm{Laser}-E_\mathrm{ex} = \gamma_{\mathrm{ex}}$ from the exciton. The FOM reaches $3\times 10^{-3}\,\pi\,\mathrm{meV}^{-1}\,\mu\mathrm{m}^{-1}$ for the lower polariton branches, so that a blueshift of $5$~meV is sufficient for a $\pi$ phase shift after $67\,\mu\mathrm{m}$ of propagation. For the upper polariton branches, the FOM can reach $14\times 10^{-3}\,\pi\,\mathrm{meV}^{-1}\,\mu\mathrm{m}^{-1}$, reducing the required propagation length to $14\,\mu\mathrm{m}$ for the same blueshift.

Active-material position and thickness, as well as the overall waveguide thickness, provide additional free parameters for engineering a dispersion with even larger differences in wavevector sensitivity between the two polariton modes, thus further optimizing the FOM. Nevertheless, the energy shifts and propagation distances required by the present proposal are already within experimental capabilities~\cite{Fieramosca2025}.
Comparable switching lengths down to 20~$\mu$m have been reported in slow-light Mach-Zehnder interferometers and bimodal waveguides in silicon photonics~\cite{Torrijos-Moran2021, Camargo2006}, but electro-optical and thermo-optical modulation in those platforms limits the achievable switching speed. All-optical switching can overcome this limitation with $\pi$ phase shifts having been demonstrated in a comparably-sized polaritonic Mach-Zehnder geometry with potential switching speeds of $\sim$100~ps in a GaAs platform~\cite{Sturm2014}. However, that approach requires spatially localized optical pumping of one of the arms of the interferometer. The multimode waveguide geometry proposed here relaxes this constraint, since both polariton modes are co-excited by a single spatially uniform pump, eliminating the need for strict alignment of the excitation.

\subsection{Continuous Real-Space Field Reshaping and Modal Switching in the SSC Regime} \label{Sec_Applications_Switch}

Having demonstrated phase control in the multimode strong coupling regime, we now turn to the SSC regime, where the avoided-crossing topology enables qualitatively different modes of operation. Near the crossover point of the mid-thickness waveguide ($D = 400$~nm, $\Delta z/D = 0.3$), well into the SSC regime, polariton modes contain significant contributions from both $\mathrm{TE}_0$ and $\mathrm{TE}_1$ modes. This coherent superposition state would form even under incoherent illumination, as the coherence is inherited from the mutual coupling to the exciton. The resulting electromagnetic state can be written as:
\begin{equation}
	\ket{\psi} = \cos\theta\, \ket{\mathrm{TE}_0} + e^{i\phi}\,\sin\theta\, \ket{\mathrm{TE}_1}.
\end{equation}
\begin{figure*}[ht]
	\centering
	\includegraphics[]{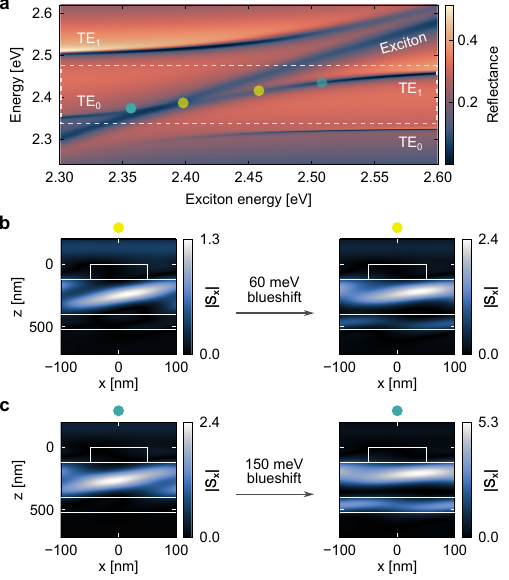}
	\caption{Exciton-tunable mode switching for the mid-thickness waveguide ($D = 400$~nm, $\Delta z/D = 0.3$). For this application, the excitation is fixed at an incidence angle of $22^\circ$ and would cover the energy range of the two inner polaritons (gray dashed rectangle), ensuring that for any given exciton energy only one polariton mode is excited. (a)~Polariton energy in reflection as the exciton energy is varied. (b,c)~Magnitude of the $x$~component of the Poynting vector at the selected yellow (b)~and green (c)~points shown in (a), illustrating the crossover from $\mathrm{TE}_0$ to $\mathrm{TE}_1$ achievable with different amounts of exciton blueshift. All evaluated quantities of this waveguide are listed in \cref{Tab_WaveguideComparison}.}
	\label{Fig5}
\end{figure*}
By adjusting the exciton energy, one obtains continuous control over the amplitude ratio of the two modal components, setting the angles $\theta$ and $\phi$. Crucially, because $\ket{\mathrm{TE}_0}$ and $\ket{\mathrm{TE}_1}$ have distinct transverse field profiles, this continuous tuning of $\theta$ translates directly into a continuous reshaping of the electromagnetic mode profile in real space, enabling sub-wavelength control of the field concentration across the active layer. The device therefore acts as a waveguide-integrated beamsplitter in the modal degree of freedom and, equivalently, as a continuously tunable real-space field-shaping element, opening the possibility of switches based not only on frequency or output channel but also on the spatial concentration of the field itself. In the single-photon regime, the structure would prepare arbitrary qubit states encoded in the transverse mode of the waveguide, providing a compact alternative to path encoding~\cite{Wang2020,Chen2021b} that requires only a single physical waveguide rather than two spatially separated paths. Combined with mode-dependent output couplers, this enables tunable on-chip state preparation and measurement within a simple planar geometry.
The modal-switching limit of this continuous reshaping is illustrated in~\cref{Fig5}, where the SSC regime enables a continuous crossover between $\mathrm{TE}_0$-like and $\mathrm{TE}_1$-like polaritons as a function of the exciton energy at a fixed incidence angle and fixed excitation energy interval.

While the required energy shifts of 60-150~meV are comparatively high in our example, further structural optimization could reduce the requirements, such that commonly achievable exciton energy shifts of tens of meV would suffice, e.g. through nonlinear effects (phase-space filling, exciton-exciton interactions, bandgap renormalization, optical Stark effect)~\cite{Shrivastava2022}, temperature variation~\cite{Guarneri2024}, or electrostatic gating of the excitonic material~\cite{Yu2017}. Such improvements can be achieved by further increasing the waveguide thickness, which reduces the mode energy spacing and thereby lowers the required exciton energy shift. However, we chose the mid-thickness waveguide for this illustration because it offers the best reflection contrast at the given grating thickness, whereas thicker waveguides require a redesigned grating with stronger far-field coupling to maintain sufficient contrast. Overall, the presented setup effectively realizes a waveguide-integrated mode switch, routing light between two distinct output channels and offering a polaritonic alternative to nonlinear directional couplers.

\section{Conclusion}

In this work, we introduced planar waveguides as a platform for multimode polariton dispersion engineering, spanning the regime from multimode strong coupling to superstrong coupling between an exciton and multiple photonic modes in the visible spectral range. The SSC endpoint of this range is characterized by polaritons with multiple photonic components and a characteristic S-shaped dispersion. The weaker multimode strong coupling regime, accessed at smaller waveguide thickness, retains a single-photonic-mode character per polariton branch but provides the differential branch sensitivity exploited in our phase-modulation scheme.

By applying a generalized Hopfield model~\cite{Cortese2023} to both ideal waveguide modes and RCWA simulation results, we showed that the optimal strategy for maximizing the coupling strength at a given overlap is to place the active material at the upper or lower boundary of the waveguide core. An active material fraction of $\Delta z / D = 0.3$ was found to provide a good compromise between substantial coupling strength and a high overlap of $\eta_{jk} \approx 0.99$.

We further demonstrated that waveguides offer a key advantage over microcavities in that the energy spacing between photonic modes can be freely tuned by adjusting the waveguide thickness without significantly affecting the fundamental mode energy. We investigated three waveguide thicknesses ($D = 200$, $400$, and $600$~nm), spanning the range from multimode strong coupling to SSC ($g_j/\Delta E_{\mathrm{ph},12}$ up to $0.74$). This tunability enables coupling of the two lowest-order modes ($\mathrm{TE}_0$ and $\mathrm{TE}_1$), which is not generally feasible in cavities due to their large mode spacing. Coupling of these low-order modes leads to up to $44\%$ higher coupling strengths for a given overlap compared to higher-order mode pairs ($\mathrm{TE}_3$/$\mathrm{TE}_4$), while also providing more distinct field profiles.

All simulations were performed using realistic parameters for two-dimensional hybrid organic-inorganic perovskites (e.g., F-PEA), which are compatible with the proposed layered structure and represent a promising candidate for experimental realization. The waveguide approach can be extended to other excitonic materials (TMDs~\cite{Lee2023}, organic dyes~\cite{Ellenbogen2011}), to coupling between modes of different polarization or propagation loss, or to hybrid plasmonic-photonic systems~\cite{Bisht2019}.

While further optimization of material parameters, waveguide geometry, and fabrication strategy is needed, the requirements are well within reach of current experimental capabilities and are based on established fabrication techniques~\cite{Chang2023, Roh2021, Fieramosca2025, Glebov2025}. Compared to cavities and metasurfaces, the in-plane propagation of waveguide modes offers the advantage of being directly compatible with existing on-chip photonic components. This positions planar waveguides as a promising route toward integrated photonic applications, such as ultrafast mode switching driven by transient exciton energy shifts and the generation of coherent modal superposition states for quantum photonic circuits. Overall, these results establish planar waveguides as a versatile, scalable platform for translating superstrong coupling physics from fundamental research into practical integrated photonic devices at visible wavelengths and room temperature.

\section{Methods}

\subsection{Waveguide Materials and Design} \label{App_Design}

As parameters for the active material layer in the waveguide, we choose values comparable to common two-dimensional hybrid organic-inorganic perovskite crystals~\cite{Song2021} with a constant background refractive index of $2.3$ ($\epsilon_\infty = 2.3^2$), a single Lorentzian oscillator with a resonance at $517$~nm ($E_\mathrm{ex} = 2.398$~eV), an oscillator strength corresponding to $\hbar \omega_p = 0.5$~eV, and a FWHM of $\gamma_{\mathrm{ex}} = 30$~meV (see definition for Lorentzian material in Section~S2 of the Supporting Information). 
Recent experiments have shown that it is possible to grow uniform single crystals of 4-Fluorophenethylammonium lead iodide (F-PEA) perovskite with the required thicknesses and optical properties~\cite{Fieramosca2018a, Coriolano2021, Fieramosca2019} making it a promising candidate for experimental realization of the proposed structures. Another suitable material candidate among 2D perovskites is phenylethyl ammonium lead iodide (PEPI), for which strong coupling has been repeatedly demonstrated \cite{Dang2020, Asham2022a, Al-Ani2023b}.

The remainder of the core is filled with a lossless dielectric that matches the background refractive index of the active material ($n_0 = 2.3$). This can be achieved experimentally by depositing a TiO$_2$ layer via electron-beam evaporation. On top of the waveguide, we define a 125~nm thick grating layer composed of the same lossless dielectric, with a filling factor of 0.5. This grating enables efficient coupling of light into the waveguide modes from above. Experimentally, the grating layer can be fabricated by electron-beam lithography (EBL), which minimizes damage to the perovskite crystal, as it is protected by the TiO$_2$ layer during this step.

The thickness of the grating is adjusted to tune the Q-factor of the photonic modes. For optimal coupling, it must be thick enough to ensure strong interaction (i.e., lowering the photonic Q-factor towards that of the exciton), yet thin enough to avoid significant alteration of the mode profiles. The grating periodicity is set to $P = 200$-$210$~nm, enabling coupling to the TE$_0$ and TE$_1$ modes at experimentally accessible angles of incidence. The waveguide is cladded below by a semi-infinite SiO$_2$ substrate with a constant refractive index of $1.5$.

\subsection{Explicit Form of Hamiltonian} \label{App_Theory}

In the case of two photonic modes ($N=2$), the overlap matrix is determined by a single parameter $\eta = \eta_{12}$, which determines the expansion coefficient and coupling strength matrix, and the Hamiltonian (\cref{Hamiltonian}) as follows:
\begin{align}
	\alpha_{11} &= 1, & g_{11} &= g_1, \\
	\alpha_{21} &= \eta, & g_{21} &= g_2 \, \eta, \\
	\alpha_{22} &= \sqrt{1-\eta^2}, & g_{22} &= g_2 \sqrt{1-\eta^2},
\end{align}

\begin{align}
	H = \begin{pmatrix}
		E_{\mathrm{ph},1} & 0  & g_{11} & 0 \\
		0 & E_{\mathrm{ph},2} & g_{21} & g_{22} \\
		g_{11} & g_{21} & E_{\mathrm{ex}} & 0 \\
		0 & g_{22} & 0 & E_{\mathrm{ex}} 
	\end{pmatrix}
\end{align}

Note that for general $N$,~\cref{Hamiltonian_DefiningEq} must be used to invert the relation between $\alpha_{jk}$ and $\eta_{jk}$, with $\alpha_{jk} = 0$ for $k > j$ and $\alpha_{11}$ real. These constraints can always be applied and reflect the choice of a particular basis for the degenerate exciton states.

\subsection{Waveguide Mode Energy Spacing} \label{App_WaveguideModeSpacing}

The energy of waveguide modes $E_j$ is determined by the effective refractive index $n_{\mathrm{eff},j}$, which depends on the waveguide thickness, the extent of the evanescent field into the surrounding medium, and the mode order. In the limit of negligible evanescent fields, the mode energy can be approximated as a function of the in-plane momentum $\beta = k_0 n_{\mathrm{eff}}$ (the propagation constant) as:

\begin{equation}
E_j(\beta) = \frac{\hbar c}{n_0} \sqrt{\beta^2 + k_{\perp,j}^2}
\end{equation}
where $k_{\perp,j} = j\pi n_0 / D$ is the out-of-plane momentum of the $j$-th mode ($\mathrm{TE}_{j-1}$) due to confinement along the $z$-direction, $k_0$ is the free-space wavevector, and $n_0$ is the refractive index of the waveguide material. In the limit where the waveguide is much thicker than the wavelength, a Taylor expansion yields:
\begin{equation}
E_j(\beta) \approx \frac{\hbar c}{n_0} \beta \left[1 + \underbrace{\frac{\pi^2 n_0^2}{2 \beta^2} \frac{j^2}{D^2}}_{\ll 1}\right]
\end{equation}

\subsection{Comparison of Simulated Waveguide Structures} 
\begin{table}[ht]
	\centering
	\footnotesize
	\begin{tabular}{@{\hspace{1mm}}lccccccccccc@{\hspace{1mm}}}
		\toprule
		D & P & $g_j$ & $\Delta E_{\mathrm{ph},12}$ & $g_j/\Delta E_{\mathrm{ph},12}$ & $\eta_{12}$ & $\Delta E_{\mathrm{pol},jk}$ & $\gamma_{\mathrm{ex}}$ & $\gamma_{\mathrm{TE}_0}$ & $\gamma_{\mathrm{TE}_1}$ & SC$_j$ & Fit RMS \\
		{[nm]} & {[nm]} & {[meV]} & {[meV]} & {} & {} & {[meV]} & {[meV]} & {[meV]} & {[meV]} &  & {[meV]} \\
		\midrule
		200 & 210 & 44-49 & 391 & 0.11-0.13 & 0.990 & 3.0 & 30 & 8.5 & 29.7 & 2.00 - 1.64 & 3.6 \\
		400 & 200 & 40-58 & 152 & 0.26-0.38 & 0.972 & 13.8 & 30 & 1.6 & 4.2 & 1.88 - 2.71 &4.2 \\
		600 & 200 & 37-59 & 80 & 0.46-0.74 & 0.995 & 2.6-10.1 & 30 & 0.9  & 1.9 & 1.74 - 2.78 & 2.7 \\
		\bottomrule
	\end{tabular}
	\caption{Comparison of simulated waveguide structures for an active-material fraction $\Delta z/D=0.3$. Shown are the waveguide thickness $D$, grating period $P$, $\mathrm{TE}_0$-$\mathrm{TE}_1$ energy spacing $\Delta E_{\mathrm{ph},12}$, coupling strength $g_j$ (first value for $\mathrm{TE}_0$, second for $\mathrm{TE}_1$), modal overlap $\eta_{12}$, fit root-mean-square error per point, full-width at half-maximum linewidths $\gamma_{\mathrm{ex}}$ (exciton) and $\gamma_{\mathrm{ph},j}$ (photonic modes), the polariton energy gap at crossover $\Delta E_{\mathrm{pol},jk}$ (two values indicate two gaps exist), and the strong coupling criterion for the two photonic modes SC$_j = g_j / \sqrt{(\gamma_{\mathrm{ex}}^2 + \gamma_{\mathrm{TE}_{j-1}}^2)/2}$ \cite{Zhang2018} (first value for $\mathrm{TE}_0$, second for $\mathrm{TE}_1$).}
	\label{Tab_WaveguideComparison}
\end{table}

\section*{Funding Sources}

EC acknowledges financial support from the Engineering and Physical Sciences Research Council (EPSRC), grant No. EP/Y021835/1. 
JB, SDM, SDL, and DB acknowledge financial support under the National Recovery and Resilience Plan (NRRP), Mission 4, Component 2, Investment 1.1, Call for tender No.~104 published on 02/02/2022 by the Italian Ministry of University and Research (MUR) under the PRIN grant 2022, funded by the European Union – NextGenerationEU – Project Title PENNA - CUP B53D23003790006 - Grant Assignment Decree No.~957 adopted on 30/06/2023 by the Italian Ministry of University and Research (MUR).

\section*{Associated Content}

\subsection*{Supporting Information}

The following files are available free of charge.
\begin{itemize}
  \item Supporting Information: Overview of works on SSC, details on simulations, additional simulation results for 200 nm, 400 nm, and 600 nm thick waveguides, theoretical model for overlap and coupling strength for ideal waveguide modes, theoretical model for interference-based mode switching (PDF)
\end{itemize}

\printbibliography

\end{document}


\maketitle
\begin{center}
	20 pages \textbar{} 17 figures \textbar{} 1 table
\end{center}
Here, we report additional results and relevant background information on the following topics:

\tableofcontents

\clearpage
\begin{landscape}
\section{Overview of superstrong coupling regime in different systems}

\begin{table}[H]
	\hspace*{-.2cm}
	\centering
	\scriptsize
	\begin{tabular}{@{} l l l l l l l l l l l @{} }
		\toprule
		\textbf{Energy} & \textbf{Temp.} & \textbf{Matter} & \textbf{Photonic} & \textbf{Mode} & \textbf{Mode spacing} & \textbf{Coupling} & \textbf{SSC ratio} & \textbf{Active} & \textbf{Paper analysis includes...} & \textbf{Ref.} \\
		& & \textbf{excitation}  & \textbf{components} & \textbf{orders} & \textbf{$\bm{\Delta E_{\mathrm{ph},jk}}$} & $\bm{g_j}$ & \textbf{\bm{$g_j / \Delta E_{\mathrm{ph},jk}$}} & \textbf{fraction} & & \\
		\midrule

			\rowcolor{viswork} VIS & RT & Perovskite & Planar waveguide & 1-2 & 391 meV & 44-49 meV & 0.11-0.13 & 30\% & Overlap {\color{green}\cmark}, Hopfield coefficients {\color{green}\cmark} & 2026  \\
			\rowcolor{viswork} (517 nm) &  & exciton & (200 nm thick) & & & & &  & SSC {\color{green}\cmark}, S-shape in theory {\color{orange}\cmark} & This   \\

			\rowcolor{viswork}  &  &  & (400 nm thick) &  & 152 meV & 40-58 meV & 0.26-0.38 &  & & work  \\
			\rowcolor{viswork} & &  &  & & & & &  & &  (theory) \\

			\rowcolor{viswork}  &  &  & (600 nm thick) &  & 80 meV & 37-59 meV & 0.46-0.74& & & \\
			\rowcolor{viswork}  & &  &  & & & & &  &  &   \\
			\midrule

			\rowcolor{vislight} VIS & RT & Perovskite & Tunable  & $\sim$5-6 & $\sim$360 meV & 78 meV & 0.22 & $\sim$15\% & Overlap {\color{red}\xmark}, Hopfield coefficients {\color{red}\xmark} & 2025~\cite{Adl2025} \\
			\rowcolor{vislight} (606 nm) & & exciton & cavity & & & & & & SSC {\color{orange}\cmark}, S-shape in theory {\color{orange}\cmark}  & \\
			\midrule

			\rowcolor{vislight} VIS & RT & Perovskite & Microcavity & 6-8 & - & - & - & 18\% & Overlap {\color{green}\cmark}, Hopfield coefficients {\color{red}\xmark} & 2023~\cite{Mandal2023} \\
			\rowcolor{vislight} (605 nm) & &  exciton &  & & & & && SSC {\color{red}\xmark}, S-shape {\color{green}\cmark}  & \\
			\midrule

			\rowcolor{vislight} VIS & RT & J-aggregate & Microcavity & 11-12 & 190 meV & 48 meV & 0.25 & $\sim$15\% & Overlap {\color{red}\xmark}, Hopfield coefficients  {\color{green}\cmark} & 2021~\cite{Georgiou2021} \\
			\rowcolor{vislight} (588 nm) & &  dye exciton & & & & & & & SSC  {\color{green}\cmark}, S-shape only for low $g_j${\color{orange}\cmark} & \\

			\midrule
			\rowcolor{vislight} VIS & RT & J-aggregate  & Microcavity & 3-4 & $\sim$600 meV & 64 meV & 0.11 & 100\% & Hopfield coefficients {\color{orange}\cmark}& 2023~\cite{Godsi2023} \\
			\rowcolor{vislight} (590 nm) & & dye exciton &  & 8-9 & $\sim$200 meV & 64 meV & 0.32 & 100\% & SSC theory (100\% active fraction) {\color{green}\cmark}  & \\
			\rowcolor{vislight} & &  &  &  &  &  & &  &  S-shape only for low $g_j$ {\color{orange}\cmark} & \\			 
			\midrule

			\rowcolor{nirlight} NIR & Cold  & Cs ensemble & Fiber ring  & $\sim 5 \times 10^7$ & 7.1 MHz & 8.7 MHz & 1.23 & low &  Overlap {\color{red}\xmark}, Hopfield coefficients {\color{red}\xmark}& 2019~\cite{Johnson2019} \\
			\rowcolor{nirlight} (852 nm) & atoms & & resonator & & & & & & SSC {\color{green}\cmark}, S-shape {\color{red}\xmark} & \\
			\midrule

			\rowcolor{thzlight} THz & Cryo & GaAs QWs  & 3D photonic  & 1-2 & $\sim$44 GHz & $\sim$72 GHz & 1.64 & 0.09 \% &  Overlap {\color{green}\cmark}, Intermode correlations {\color{green}\cmark} & 2025~\cite{Tay2025} \\
			\rowcolor{thzlight} (360 GHz) & & cyclotron resonance & crystal cavity & & & & & & USC+SSC {\color{green}\cmark}, S-shape {\color{green}\cmark} & \\
			\midrule

			\rowcolor{thzlight} THz & Cryo & GaAs QWs  & patterned  & 1-2 & 0.8 THz & 0.22-0.43 THz & 0.28-0.54 & low &  Overlap {\color{green}\cmark}, Hopfield coefficients {\color{red}\xmark}& 2024~\cite{Mornhinweg2024a} \\
			\rowcolor{thzlight} (1.2 THz) & & cyclotron resonance & metasurface & & & & & & USC+SSC {\color{green}\cmark}, S-shape {\color{green}\cmark} & \\
			\midrule

			\rowcolor{mwlight} Microwave & Cryo & Transmon & Superconducting  & $\sim 250$ & $\sim$50 MHz & $\sim$65 MHz & 1.30 & low &  Overlap {\color{red}\xmark}, Hopfield coefficients {\color{red}\xmark}& 2019~\cite{Kuzmin2019a} \\
			\rowcolor{mwlight} (5-13 GHz) & &  qubit & cavity & & & & & &  SSC {\color{green}\cmark}, S-shape (very clean) {\color{green}\cmark} & \\

			\midrule
			\rowcolor{mwlight} Microwave & Cryo & Superconducting &  Acoustic  & $\sim$1000 & 9.2 MHz & 2.4 MHz & 0.26 & - &  Overlap {\color{red}\xmark}, Hopfield coefficients {\color{red}\xmark} & 2016~\cite{Han2016} \\
			\rowcolor{mwlight} (10 GHz) & &  resonator & resonator & & & & & & SSC {\color{green}\cmark}, S-shape {\color{green}\cmark} & \\

		\bottomrule
	\end{tabular}
	\caption{Summary of selected experimental works on superstrong coupling compared to our work. "Active fraction" is the ratio of active material volume to total mode volume. If references did not report certain values they were estimated from their data (marked by \mbox{$\sim$)} or omitted (\mbox{marked~"-")}. Aspects analyzed in the papers are classified as: \textcolor{green}{\cmark} = included in paper, \textcolor{orange}{\cmark} = partially included \textcolor{red}{\xmark} = not included. RT: room temperature, Cryo:~cryogenic temperatures.}
	\label{SI_Tab_Overview}
\end{table}
\end{landscape}
\clearpage

\section{Optical properties of materials used in the simulation} \label{SI_LorentzianOsc}

The permittivity $\epsilon(\omega)$ of the active material is described by a Lorentzian model:

\begin{align}
	\epsilon(\omega) = \epsilon_\infty + \frac{\omega_p^2}{\omega_0^2 - \omega^2 - \mathrm{i} \gamma_{\mathrm{ex}} \omega}
\end{align}

As reported in the main text, the parameters used in the simulations are:
\begin{itemize}
	\item[] $\epsilon_\infty = 2.3^2$ \hspace{1em} $\hbar \omega_0 = 2.398$ eV \hspace{1em} $\hbar \omega_\mathrm{p} = 0.5$ eV \hspace{1em} $\hbar \gamma_{\mathrm{ex}} = 30$ meV
\end{itemize}

With this definition, $\gamma_{\mathrm{ex}}$ corresponds to the full width at half maximum (FWHM) of $\mathrm{Im}(\epsilon)$. The dielectric filler is assumed to be lossless, with its permittivity matched to the background value of the active material $\epsilon_\infty$. The substrate is assumed to be a semi-infinite lossless SiO$_2$ layer with a constant refractive index of 1.5.

\section{Details on simulations and fitting} \label{SI_DetailsFitting}


The excitation is a plane wave incident at an angle $\theta_{\mathrm{inc}}$ with respect to the surface normal ($z$ axis) of the waveguide. The incident wavevector lies in the $xz$ plane, resulting in the electric field being polarized along the $y$ axis. During the testing phase, 20 Fourier orders were used, while 40 were used for the final plots included in this publication. The reported reflectance values are defined as the ratio of reflected to incident power.

The corresponding parameter settings in S$^4$ are (where $\theta_{\mathrm{inc}}$ corresponds to \texttt{phi} in the S$^4$ naming convention):

\begin{verbatim}
IncidenceAngles = (phi, 0)
sAmplitude = 1
pAmplitude = 0
NumBasis = 40
\end{verbatim}

To fit the theoretical model, reflection dips or peaks are identified from the second derivative of the reflectance spectrum with respect to energy.

\begin{figure*}[ht]
    \includegraphics[]{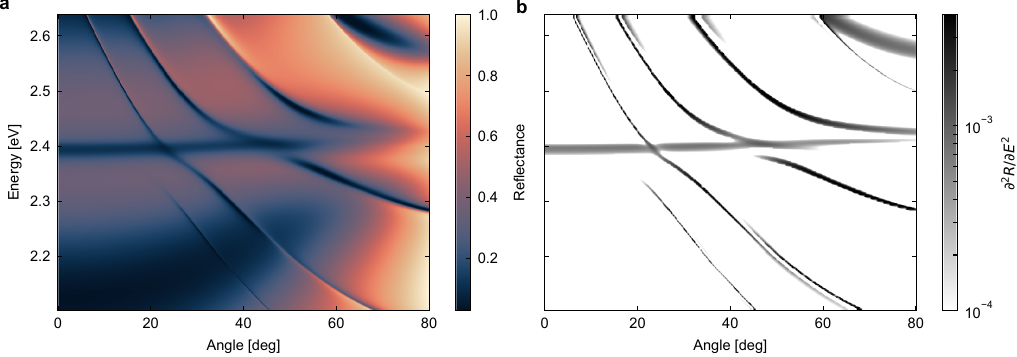}
\caption{Comparison between reflectance spectrum and its second derivative for the mid-thickness waveguide ($400$~nm) presented in~\cref{Fig3_2}. (a)~Original reflectance spectrum. (b)~Second derivative of the reflectance spectrum. The dips in the reflectance spectrum correspond to positive peaks in the second derivative. Negative values of the second derivative are rendered as white for clarity. Note that the narrow mode visible on the top right at ($2.5$~eV, $80\degree$) is not a waveguide mode and does not couple significantly to the exciton. It is therefore omitted from the analysis.}
    \label{FigS_SecondDerivative}
\end{figure*}

The dispersion of the uncoupled photonic modes is extracted by setting the oscillator strength of the exciton to zero ($\omega_\mathrm{p} = 0$). Within the studied energy and incidence angle range, the resulting dispersion curves are accurately described by a second-order polynomial, obtained by fitting $E_{\mathrm{ph},j}(\theta_{\mathrm{inc}})$.

The extracted fit parameters are used to construct the photonic part of the Hamiltonian in~\cref{Hamiltonian} of the main text. This Hamiltonian is then fitted to the polariton dispersion from the full simulation (with nonzero oscillator strength) by minimizing the RMS error between the fit and simulated data. The fitting parameters are the coupling strengths $g_j$ and the overlap matrix $\eta_{jk}$ between the photonic modes.

\section{Additional simulation results for low-, mid-, and high-thickness waveguides}

For completeness, we provide the full fitting results that were discussed in the main text here.

\subsection{Low-thickness waveguide (200 nm)}

We note that for this specific geometry, the overlap coefficient $\eta_{12}$ could not be reliably extracted from the fit, as the RMS error depends only weakly on the overlap value. The fitting procedure yields an overlap value of $\eta_{12} = 0.883$ but leaves a large gap between the polaritons that is not present in the simulated data. This issue could be resolved by giving more weight to the data points near the crossover region.

Instead, here we resort to manually fixing the overlap value to its expected value for a $30\%$ layer at the bottom of the waveguide, which is $\eta_{12} = 0.99$. We justify this by noting that all values of $\eta_{12} < 0.97$ leave a significant gap between the modes, which is absent in the simulated data and can therefore be excluded.














\begin{figure*}[ht]
	\includegraphics[]{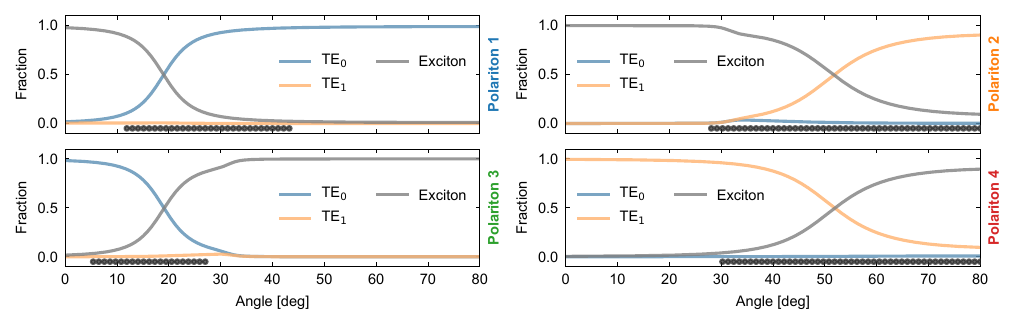}
	\caption{Fractions of uncoupled modes in the 4 polaritons of the thin waveguide ($200$~nm) shown in~\cref{Fig2}, as obtained from the fit. The crossover from $\mathrm{TE}_0$ to $\mathrm{TE}_1$ occurs between polaritons 2 and~3. Gray dots mark the angle values used for fitting each polariton. Note that the presence of dots indicates that the reflection contrast was sufficient to resolve the mode at the corresponding angle. The absence of dots may arise from insufficient contrast, the mode extending beyond the investigated energy interval, or because points where the mode is nearly purely excitonic were not included in the fit.}
	\label{FigSFractionsThin}
\end{figure*}
\clearpage

\newpage
\subsection{Mid-thickness waveguide (400 nm)}

\begin{figure*}[ht]
	\includegraphics[]{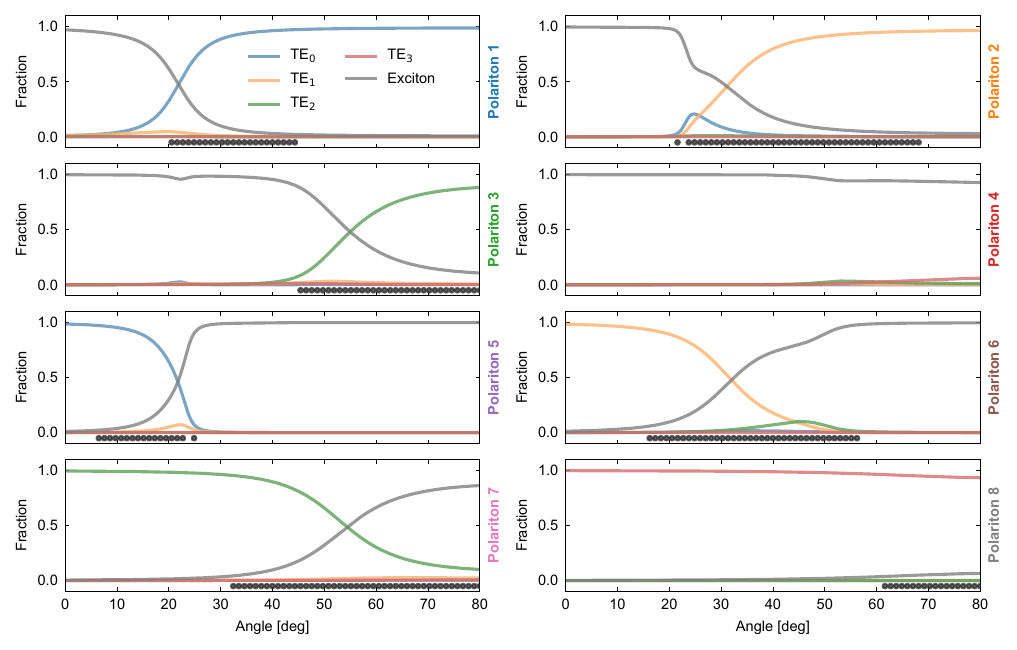}
	\caption{Fractions of uncoupled modes in the 8 polaritons of the mid-thickness waveguide ($400$~nm) shown in~\cref{Fig3_2}, as obtained from the fit. The crossover from $\mathrm{TE}_0$ to $\mathrm{TE}_1$ occurs between polaritons 5 and~2. Gray dots mark the angle values used for fitting each polariton. Note that the presence of dots indicates that the reflection contrast was sufficient to resolve the mode at the corresponding angle. The absence of dots may arise from insufficient contrast, the mode extending beyond the investigated energy interval, or because points where the mode is nearly purely excitonic were not included in the fit.}
	\label{FigSFractionsMedium}
\end{figure*}

\newpage
\subsection{High-thickness waveguide (600 nm)} \label{SI_600}

\begin{figure*}[ht]
	\includegraphics[]{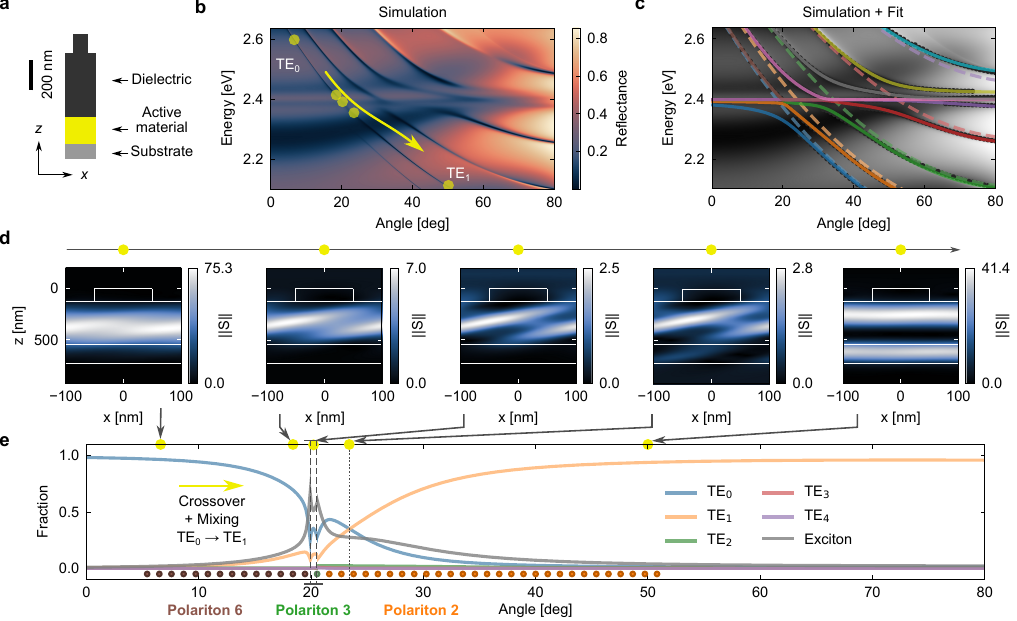}
	\caption{Simulation of superstrong coupling in the high-thickness waveguide ($D=600$~nm). (a)~Unit cell of the grating structure with active material fraction $\Delta z/D = 0.3$ and periodicity $P=200$~nm. (b)~Polariton dispersion in reflection. (c)~Fit using $\mathrm{TE}_0$-$\mathrm{TE}_4$ photonic modes (dashed lines), yielding 10 polariton branches. Gray dots indicate points used in fit. (d)~Poynting vector magnitude at selected yellow points in (b), illustrating the crossover from $\mathrm{TE}_0$ to $\mathrm{TE}_1$. (e)~Fractions of uncoupled modes in polaritons 2 (orange, right), 3 (green, middle), and 6 (brown, left) separated by the crossover lines (gray dashed). No significant discontinuity is observed due to the small energy gap. Colored dots mark angles used in the fit, their presence indicates sufficient reflection contrast.}
	\label{Fig3}
\end{figure*}

\begin{figure*}[ht]
	\includegraphics[]{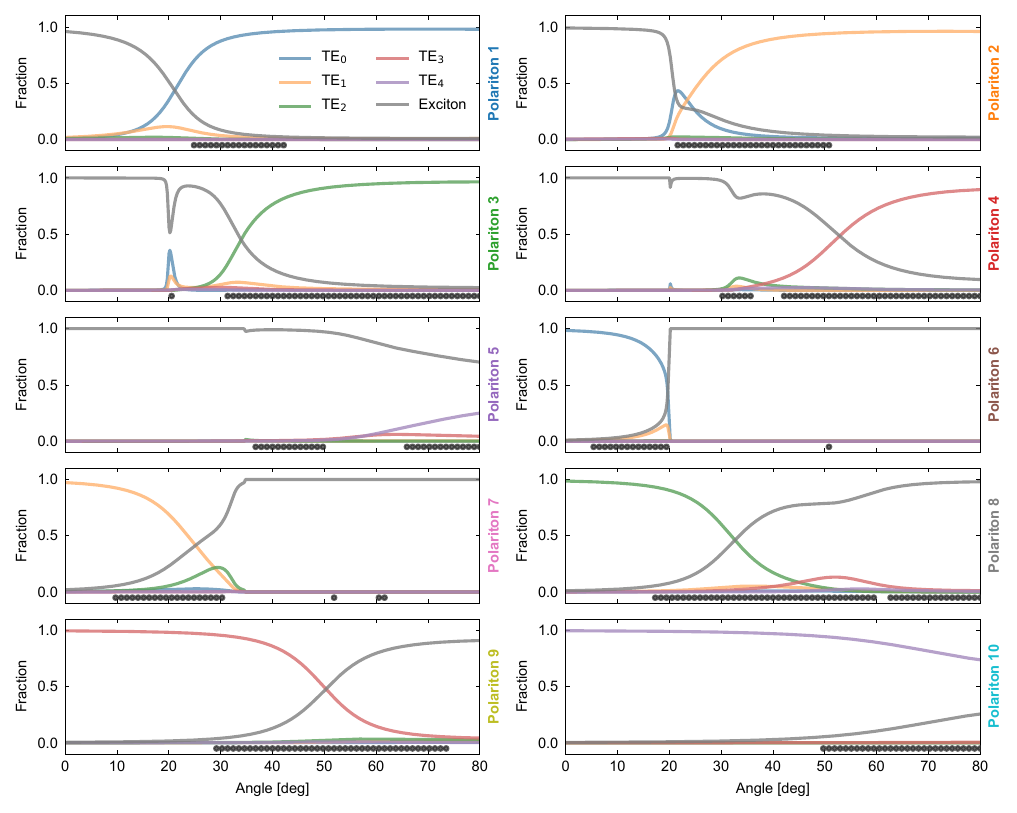}
	\caption{Fractions of uncoupled modes in the 10 polaritons of the high-thickness waveguide ($600$~nm) shown in~\cref{Fig3}, as obtained from the fit. The crossover from $\mathrm{TE}_0$ to $\mathrm{TE}_1$ occurs between polaritons 6, 3 and~2. Gray dots mark the angle values used for fitting each polariton. Note that the presence of dots indicates that the reflection contrast was sufficient to resolve the mode at the corresponding angle. The absence of dots may arise from insufficient contrast, the mode extending beyond the investigated energy interval, or because points where the mode is nearly purely excitonic were not included in the fit.}
	\label{FigSFractionsThick}
\end{figure*}

\clearpage

\newpage
\subsubsection{Sweep of active material fraction for the high-thickness waveguide}

The parameters for the grating structure are the same as in~\cref{Fig3}.

\begin{figure*}[ht]
	    \centering
		\includegraphics[height=0.68\textheight]{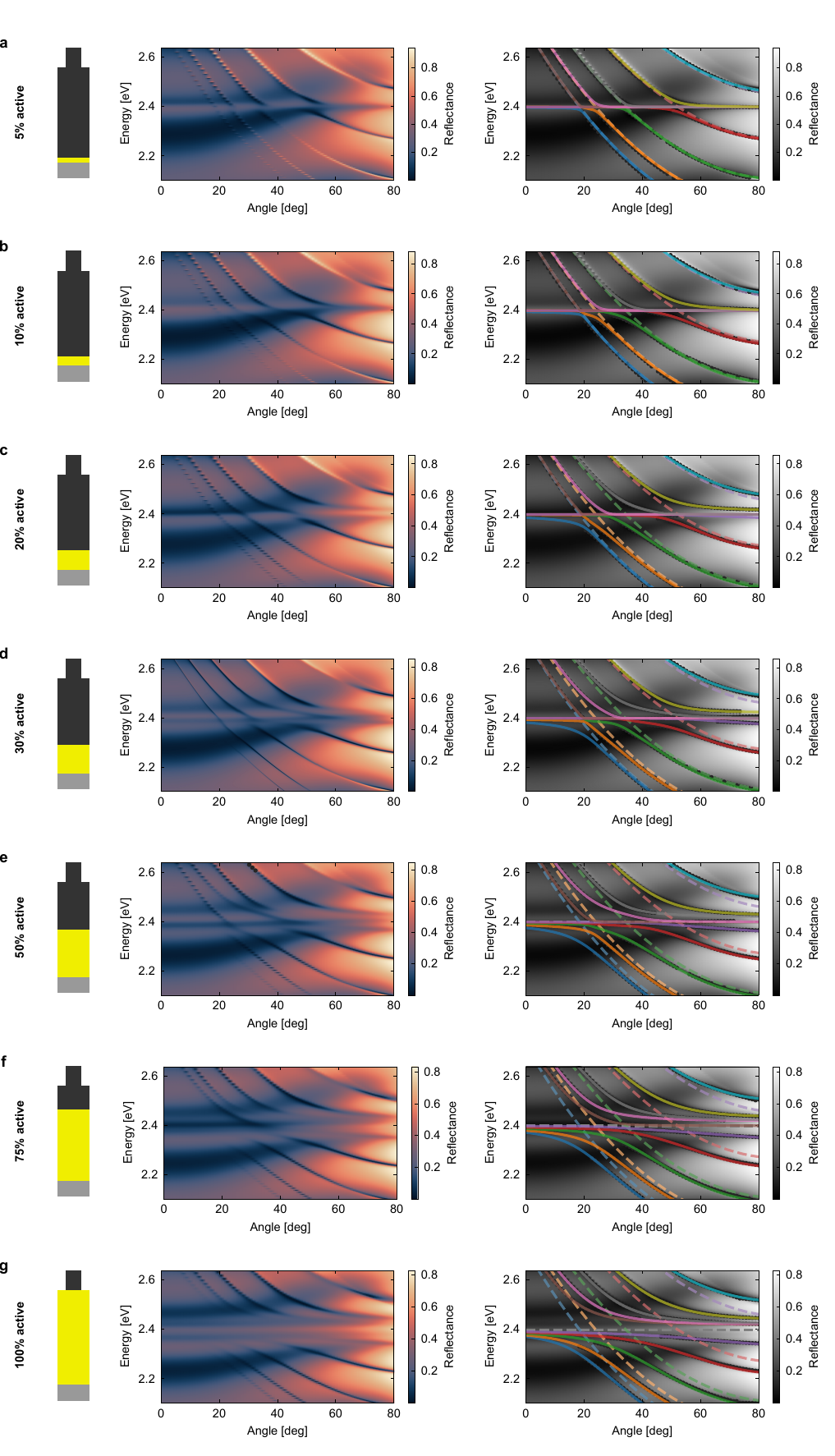}
		\caption{Simulation of high-thickness waveguide ($D=600$~nm) for different active material fractions $\Delta z/D$: 5\%~(a), 10\%~(b), 20\%~(c), 30\%~(d), 50\%~(e), 75\%~(f), and 100\%~(g). Left: Geometry of the waveguide. Middle: Polariton dispersion in reflection. Right: Fit using $\mathrm{TE}_0$-$\mathrm{TE}_4$ photonic modes (dashed lines), yielding 10 polariton branches. Gray dots indicate points used in the fit.}
	\label{FigS_ActiveFractionSweep_600nm_A}
\end{figure*}

\clearpage
\begin{figure*}[ht]
	\centering
	\includegraphics[height=0.68\textheight]{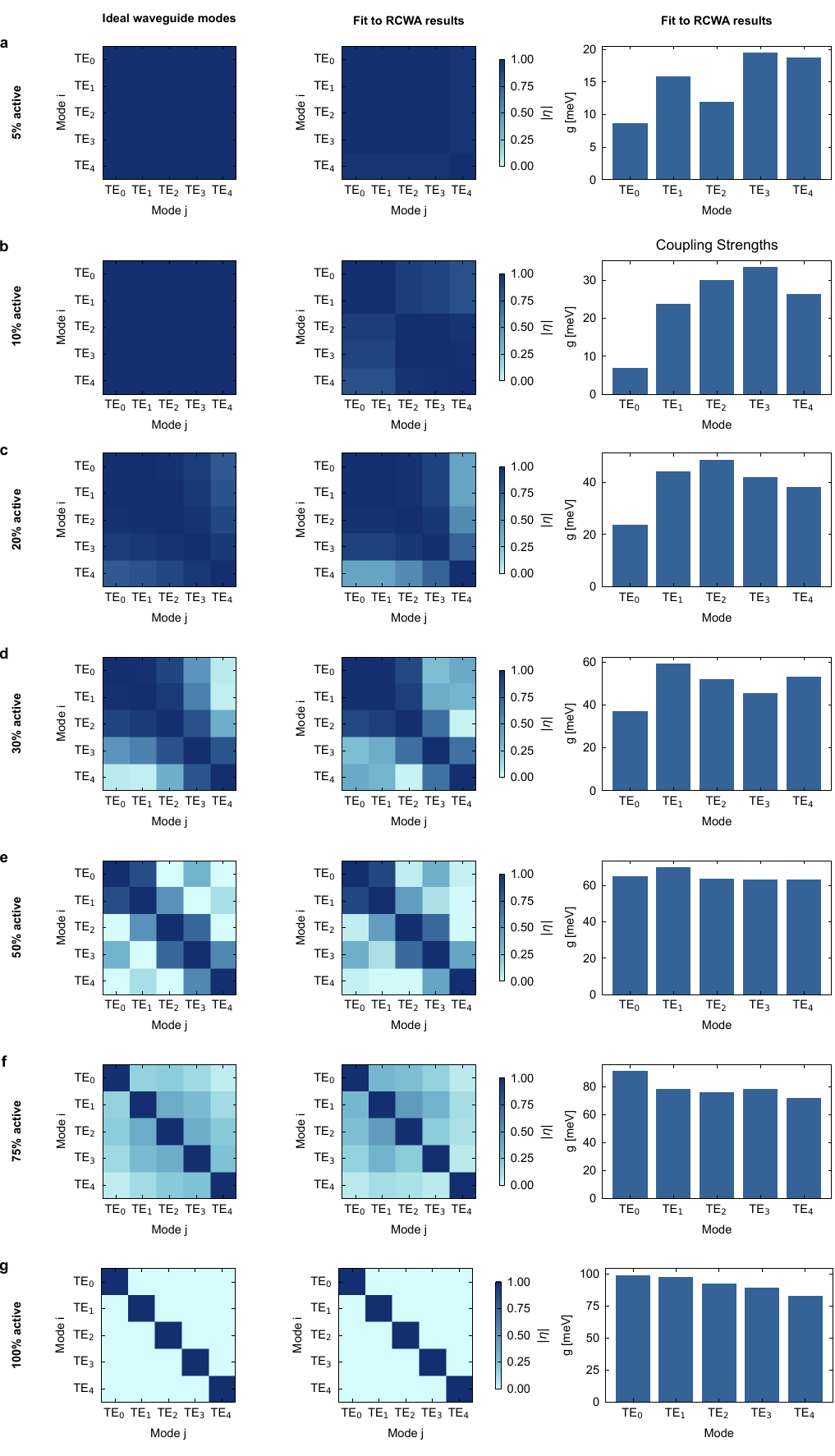}
	\caption{Continuation of~\cref{FigS_ActiveFractionSweep_600nm_A}: Simulation of high-thickness waveguide ($D=600$~nm) for different active material fractions $\Delta z/D$: 5\%~(a), 10\%~(b), 20\%~(c), 30\%~(d), 50\%~(e), 75\%~(f), and 100\%~(g). Left: Overlap matrix $\eta_{jk}$ obtained from~\cref{Eq_IdealWaveguideModel}. Middle: Overlap matrix $\eta_{jk}$ obtained from the fits shown in~\cref{FigS_ActiveFractionSweep_600nm_A}. Right: Total coupling strengths $g_j$ obtained from the fit.}
	\label{FigS_ActiveFractionSweep_600nm_B}
\end{figure*}

\clearpage
\section{Magnitude of energy gap between inner polaritons is not an indicator of mode overlap} \label{SI_EnergyGap}

In the general case of an imperfect overlap, $\eta_{jk} < 1 $, the used Hamiltonian (\cref{Hamiltonian}) of the main text predicts a finite energy gap $\Delta E_{\mathrm{pol},jk}$ between the two inner polaritons at the crossover. If this energy gap is smaller than the linewidth of the polaritons, the gap cannot be resolved and the transition appears continuous.

\begin{figure*}[ht]
	\includegraphics[]{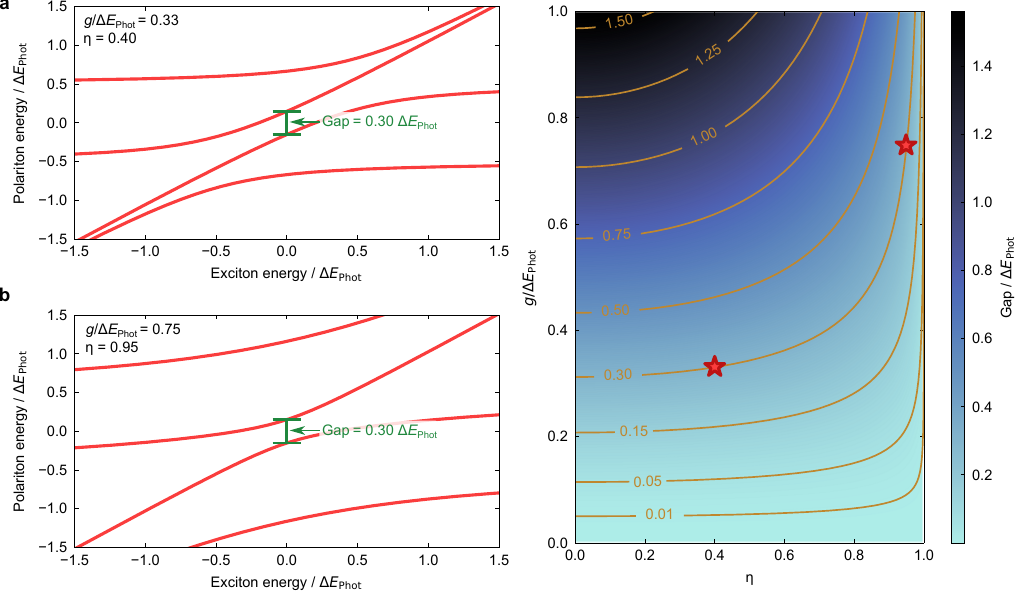}
	\caption{Energy gap between inner polaritons as a function of coupling strength and mode overlap, calculated using the Hamiltonian in~\cref{Hamiltonian} of the main text for the case of two photonic modes. (a)~Polariton energies for varying exciton energies for a combination of low overlap and low coupling strength. (b)~Polariton energies for a combination of high overlap and high coupling strength yielding the same energy gap (green marker) as in~(a). (c)~Energy gap between the inner polaritons as a function of coupling strength and mode overlap. The ochre contour lines indicate the combinations of coupling strength and overlap that yield a given gap size. All energies are normalized to the photonic mode separation $\Delta E_{\mathrm{ph},12}$, and the photonic modes are assumed to have energies of $\pm\,\Delta E_{\mathrm{ph},12}/2$.}
	\label{FigS_GapSize}
\end{figure*}

One might expect the magnitude of this energy gap to be a direct indicator of the mode overlap, with smaller gaps corresponding to higher overlaps. However, this is not the case, as the gap size also depends on the coupling strength. A small gap can result from either a high overlap or a low coupling strength. In fact, there are infinitely many combinations of overlap and coupling strength that yield the same gap size, as illustrated by the ochre contour lines in~\cref{FigS_GapSize}(c). Therefore, the mode overlap cannot be reliably inferred from the gap size alone and must instead be extracted from a full fit of the polariton dispersion.

Note that the Hamiltonian used here does not explicitly include losses as parameters. The presence or absence of a continuous crossover in the dispersions can therefore only be inferred by comparing the calculated gap size to the linewidth of the polaritons at the crossover. A different model that explicitly includes losses but is restricted to the limiting cases of zero or unity overlap, has been developed for microcavities~\cite{Balasubrahmaniyam2021}.

\clearpage
\section{Theoretical considerations on optimizing the mode overlap} \label{SI_Sec_IdealWaveguideTheory}

In order for the involved photonic modes to couple efficiently, their overlap $\eta_{jk}$ needs to be optimized which is achieved by strategically placing the active material where the modal overlap is high.

We start by assuming that the electric field profiles $F_{y,{\mathrm{TE}_0}}(x,z)$ and $F_{y,{\mathrm{TE}_1}}(x,z)$ of the two photonic modes in the presence of the active material are known. (The $y$ dependence is omitted due to the invariance of the structure in this direction.) In practice, we use the waveguide modes in the absence of the active material as an initial guess, which could later be refined iteratively once the first round of the optimization is complete.

In the case of the layered fabrication approach that we target in this work, the most general form for the overlap given a certain desired total thickness of the active material $\Delta z$ arranged in a number $n_{\mathrm{layer}}$ of layers is:

\begin{equation}
\eta_{12} = \frac{ \sum_{i=1}^{n_{\mathrm{layer}}} \int_{a_i}^{b_i} dz \int dx \, F_{y,{\mathrm{TE}_0}}^{*}(x,z) F_{y,{\mathrm{TE}_1}}(x,z)}{\sqrt{\left (  \sum_{i=1}^{n_{\mathrm{layer}}} \int_{a_i}^{b_i} dz \int dx \, |F_{y,{\mathrm{TE}_0}}(x,z)|^2\,\right ) \;\; \left(   \sum_{i=1}^{n_{\mathrm{layer}}} \int_{a_i}^{b_i} dz \int dx \, |F_{y,{\mathrm{TE}_1}}(x,z)|^2\,\right)}},
\end{equation}
where $a_i,b_i$ describe the boundaries of layer $i$ along the $z$ direction, which are constrained by $\Delta z = \sum_{i=1}^{n_{\mathrm{layer}}} b_i-a_i$, and the integral over $dx$ runs over the complete length of the unit cell of the grating.

The optimization procedure in this general case would encompass finding the optimal values of $a_i,b_i$ and the number of layers $n_{\mathrm{layer}}$ in order to maximize $\eta_{12}$. This is a numerically demanding task, the solution of which might benefit from techniques used in generalizations of the maximum subarray problem.

In our work, we are restricting ourselves to $n_{\mathrm{layer}}=1$ as higher numbers of layers would only become beneficial when targeting the coupling of modes of higher order, or when the grating strongly distorts the waveguide modes. In the case of $n_{\mathrm{layer}}=1$, the optimization procedure is simplified to maximizing:

\begin{equation} \label{SI_OverlapEq}
	\eta_{12}(z_0) = \frac{ \int_{z_0-\Delta z/2}^{z_0+\Delta z/2} dz \int dx \, F_{y,{\mathrm{TE}_0}}^{*}(x,z) F_{y,{\mathrm{TE}_1}}(x,z)}{\sqrt{\left ( \int_{z_0-\Delta z/2}^{z_0+\Delta z/2} dz \int dx \, |F_{y,{\mathrm{TE}_0}}(x,z)|^2\,\right ) \;\; \left(  \int_{z_0-\Delta z/2}^{z_0+\Delta z/2} dz \int dx \, |F_{y,{\mathrm{TE}_1}}(x,z)|^2\,\right)}},
\end{equation} 
as a function of the center position $z_0$ of the active material layer, given a fixed thickness $\Delta z$. Note that due to the terms in the denominator, this optimization procedure is different from just picking the locations where the product $F_{y,{\mathrm{TE}_0}}(x,z) F_{y,{\mathrm{TE}_1}}(x,z)$ is maximal.

 In Section~\ref{SI_Ideal}, we show the analytical solution for this optimization problem for an ideal waveguide. For the simulation results in the main text, this problem is solved numerically with a brute-force approach.

\subsection{Overlap and coupling strength optimization in ideal waveguides} \label{SI_Ideal}

As we will see, a high overlap can quite easily be achieved by using infinitesimally thin layers of the active material (e.g., TMD monolayers). However, this comes at the cost of low coupling strength as the fraction of the mode volume filled by the active material would be small. To outline this tradeoff between overlap $\eta_{12}$ and coupling strengths $g_1$ and $g_2$ between the photonic modes and the exciton in the active material, we study the problem on the example of an ideal waveguide.

\begin{figure*}[ht] 
	\includegraphics[width=170mm]{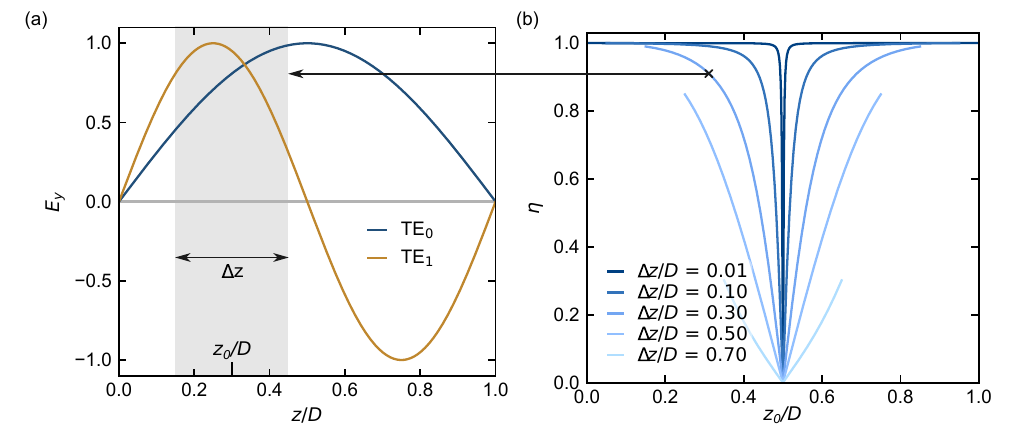}
	\caption{Dependence of mode overlap on position of active material in an ideal waveguide. (a)~Electric field of ideal TE$_0$ and TE$_1$ waveguide modes. Gray rectangle indicates an example placement of the active material with center position $z_0$ and thickness $\Delta z$ in a waveguide of total thickness $D$ (b)~Mode overlap as a function of the active material position for several values of its thickness. Curves terminate where the thickness of the material restricts the allowable positions.}
	\label{FigSIdeal1}
\end{figure*}

To this end, we assume perfectly reflecting waveguide walls and a negligibly thin grating, resulting in the following simple electric field profiles (corresponding to~\cref{Eq_IdealWaveguideModel} of the main text):
\begin{equation}
\begin{aligned}
  F_{y,{\mathrm{TE}_0}}(x,z) &= \sin\!\Big(\frac{\pi z}{D}\Big), \\
  F_{y,{\mathrm{TE}_1}}(x,z) &= \sin\!\Big(\frac{2 \pi z}{D}\Big)
\end{aligned}
\label{EQ_IdealWaveguideFields}
\end{equation}

where $D$ is the total thickness of the waveguide (\cref{FigSIdeal1}(a)). In this case, the integrals in~\cref{SI_OverlapEq} and the maximization of $\eta_{12}$ can in principle be solved analytically. However, the resulting expressions are lengthy and not particularly instructive. Therefore, we resort to numerical methods.

First, we show the dependence of the overlap on the position of the active material for a few different thicknesses $\Delta z$ of the active material (~\cref{FigSIdeal1}(b)).

As evident, the highest overlap attainable for any given thickness is obtained when the active material is placed at either the top- or bottommost position within the waveguide (maximal or minimal allowed value of $z_0$). In the limit of an infinitesimally thin active layer located at these positions, the overlap approaches unity. This behavior can be understood by noting that at the waveguide boundaries ($z/D=0$ or $z/D=1$), both electric fields vanish and may be approximated locally by the linear term of their respective Fourier expansions, albeit with different slopes. Upon normalization according to~\cref{SI_OverlapEq}, these linear field profiles become identical, and as such yield an overlap of 1 by definition.

In general, thinner materials achieve a high overlap in any location of the waveguide, except in the center where positive and negative contributions to the overlap integral compensate each other. With increasing thickness of the material, the maximally achievable overlap decreases.

\begin{figure*}[ht] 
	\includegraphics[width=170mm]{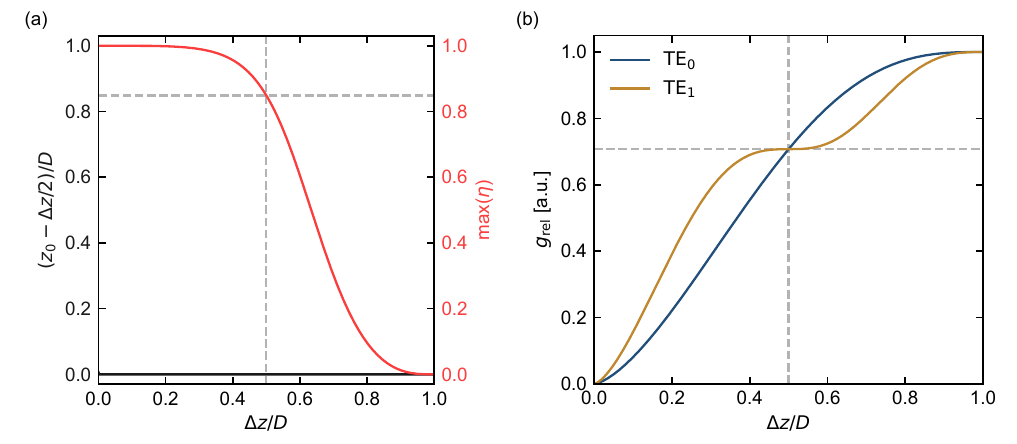}
	\caption{1D Optimization of the overlap $\eta_{12}$ for a given material thickness $\Delta z$ using ideal TE$_0$ and TE$_1$ waveguide modes. (a)~Maximally achievable overlap and the corresponding optimal position of the material. The result of the optimization places the material at the bottom of the waveguide for all thicknesses. By symmetry, an equivalent solution exists at the top (position not shown for clarity). (b)~Relative coupling strengths corresponding to the solutions in (a). $g_\mathrm{rel}=1$ corresponds to the coupling strength in a completely filled waveguide.}
	\label{FigSIdeal3}
\end{figure*}

To illustrate this point further, we numerically calculate the maxima of~\cref{SI_OverlapEq} for a given thickness of the active material, yielding the highest achievable overlap and the optimal position of the material (\cref{FigSIdeal3}). The result of the optimization confirms that the optimal position for the active material to achieve high overlap is the top- or bottommost position within the waveguide.

To quantify how strongly each photonic mode interacts with the active material in this case, we use the following relative measure of the coupling strength~\cite{Tserkezis2020}:

\begin{equation}
	g_\mathrm{rel} = \sqrt{\int_{z_0-\Delta z/2}^{z_0+\Delta z/2} dz \, |F_{y}|^2}
\end{equation}  

With this definition, we can rewrite~\cref{SI_OverlapEq} as:

\begin{equation} \label{SI_OverlapEq_g}
	\eta(z_0) = \frac{ \int_{z_0-\Delta z/2}^{z_0+\Delta z/2} dz \int dx \, F_{y,{\mathrm{TE}_0}}^{*}(x,z) F_{y,{\mathrm{TE}_1}}(x,z)}{g_\mathrm{rel,\mathrm{TE}_0} \ g_\mathrm{rel,\mathrm{TE}_1}}.
\end{equation} 

As the relative coupling strengths appear in the denominator,~\cref{SI_OverlapEq_g} directly shows the tradeoff between coupling strength and overlap. This is evident in~\cref{FigSIdeal3}(b), where the coupling strengths decrease as the optimal solution achieves higher overlap.

As a practical result of this discussion, when the lower (or upper) half of the waveguide is filled by the active material ($\Delta z = 0.5$) one can expect an overlap of the TE$_0$ and TE$_1$ modes of $\eta_{12} = 0.85$ and an equal coupling strength for both modes that reaches $1/\sqrt{2}$ of the value in the completely filled waveguide. \\

\begin{figure*}[ht] 
	\includegraphics[width=170mm]{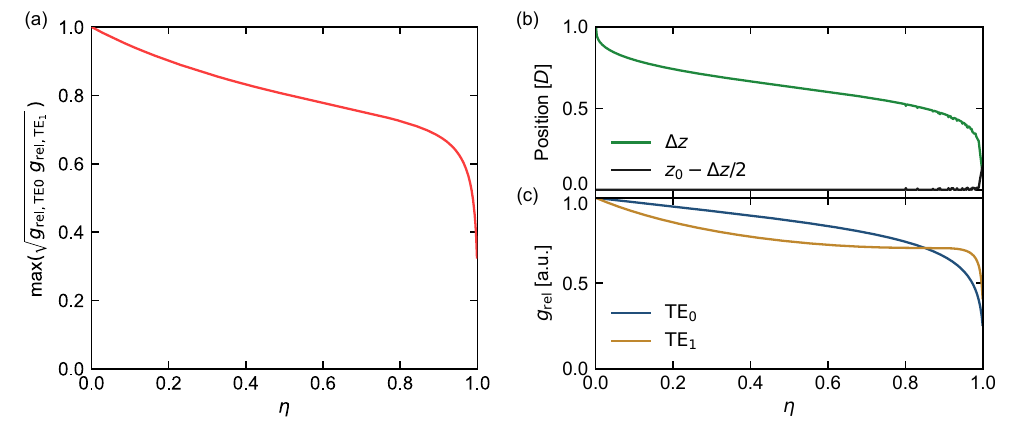}
	\caption{2D Optimization of the geometric mean of the coupling strengths for a given overlap $\eta_{12}$ using ideal TE$_0$ and TE$_1$ waveguide modes, with both material thickness $\Delta z$ and position $z_0$ as free parameters. (a)~Maximally achievable geometric mean of the coupling strengths for a given overlap. (b)~The optimal values of the position and thickness of the active material to achieve the solution in (a).~The result of the optimization places the material at the bottom of the waveguide for all thicknesses. By symmetry, an equivalent solution exists at the top (position not shown for clarity). (c)~The resulting individual coupling strengths for the TE$_0$ and TE$_1$ modes for the solution in (a). $g_\mathrm{rel}=1$ corresponds to the coupling strength in a completely filled waveguide. For large $\eta_{12}$, the low thickness causes numerical noise due to many nearly equivalent positions (cf.~\cref{FigSIdeal1}(b)).}
	\label{FigSIdeal4}
\end{figure*}

\begin{figure*}[ht] 
	\includegraphics[width=170mm]{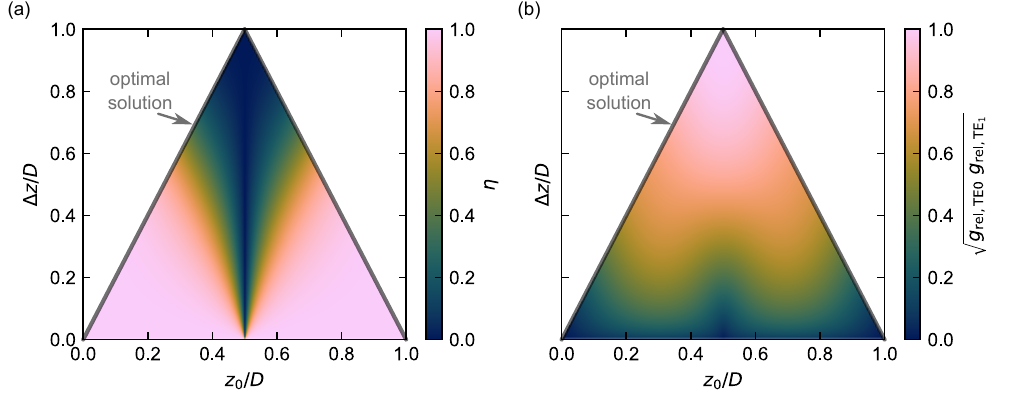}
	\caption{Dependence of (a)~the overlap $\eta_{12}$ and (b)~the geometric mean of the coupling strengths on the material thickness $\Delta z$ and position $z_0$ for ideal TE$_0$ and TE$_1$ waveguide modes. This library is used for the brute-force optimization shown in~\cref{FigSIdeal4}. The optimal solution, corresponding to placement of the active material at the top- or bottommost position of the waveguide, is highlighted by the gray line.}
	\label{FigSIdeal4b}
\end{figure*}

Although the previous optimization provided clear guidance, one might still wonder whether a better solution exists that directly maximizes the coupling strength for a given overlap, treating both $z_0$ and $\Delta z$ as free optimization parameters. To address this, we performed a 2D brute-force optimization using the geometric mean of the coupling strengths $\sqrt{g_\mathrm{rel,\mathrm{TE}_0} \ g_\mathrm{rel,\mathrm{TE}_1}}$ as the optimization target (\cref{FigSIdeal4,FigSIdeal4b}).

Again, we find that the optimal configuration places the active material at either the top- or bottommost position of the waveguide. The results in~\cref{FigSIdeal4} therefore provide a useful guideline for the design of realistic geometries, illustrating the tradeoff between coupling strength and mode overlap.

\clearpage
\subsection{Overlap and coupling strength optimization for higher-order modes} 

Although we are focusing on the TE$_0$ and TE$_1$ modes in this manuscript, we are including the situation that the simple model predicts for the TE$_3$ and TE$_4$ modes of the ideal waveguide, analogous to the treatment used before.

\begin{figure*}[ht] 
	\includegraphics[width=170mm]{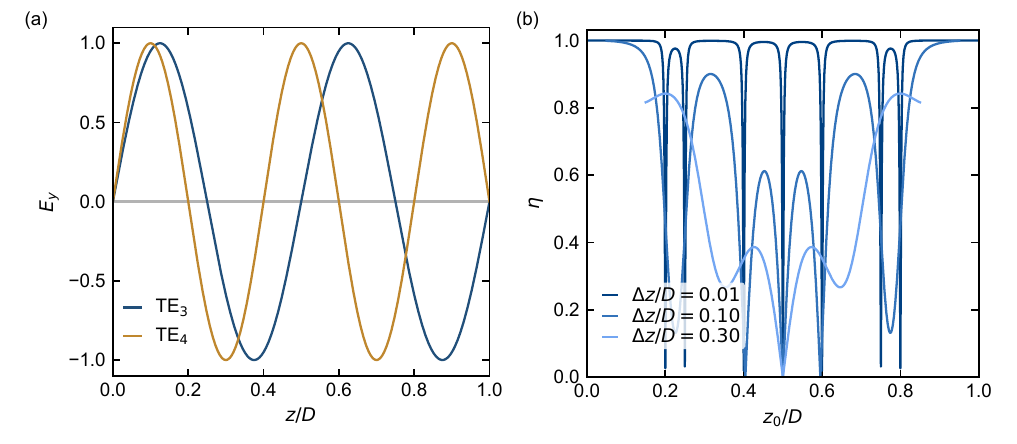}
	\caption{Dependence of mode overlap on position of active material for higher-order modes. (a)~Electric field of ideal TE$_3$ and TE$_4$ waveguide modes. (b)~Mode overlap as a function of the active material position for several values of its thickness. Curves terminate where the thickness of the material restricts the allowable positions. For thin active layers, the overlap exhibits 7 minima at positions where one of the modes crosses zero while the other remains finite.}
	\label{FigSIdeal2}
\end{figure*}

\begin{figure*}[ht] 
	\includegraphics[width=170mm]{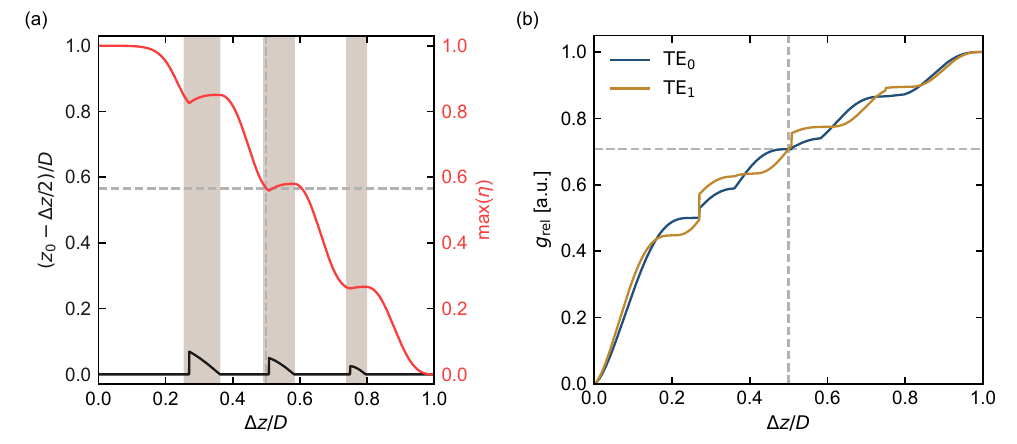}
	\caption{1D Optimization of the overlap $\eta_{45}$ for a given material thickness $\Delta z$ using ideal TE$_3$ and TE$_4$ waveguide modes. (a)~Maximally achievable overlap and the corresponding optimal position of the material. Brown shading highlights where overlap does not increase with decreasing material thickness. By symmetry, an equivalent solution exists at the top (position not shown for clarity). (b)~Relative coupling strengths corresponding to the solutions in (a). $g_\mathrm{rel}=1$ corresponds to the coupling strength in a completely filled waveguide.}
	\label{FigSIdeal3H0}
\end{figure*}

As shown in~\cref{FigSIdeal2,FigSIdeal3H0}, when the thickness of the active material is fixed (e.g., due to experimental constraints) the position yielding the highest overlap does not generally coincide with the top- or bottommost placement. Moreover, the overlap does not increase monotonically with decreasing thickness for higher-order modes, making certain thickness values unfavorable.

\begin{figure*}[ht] 
	\includegraphics[width=170mm]{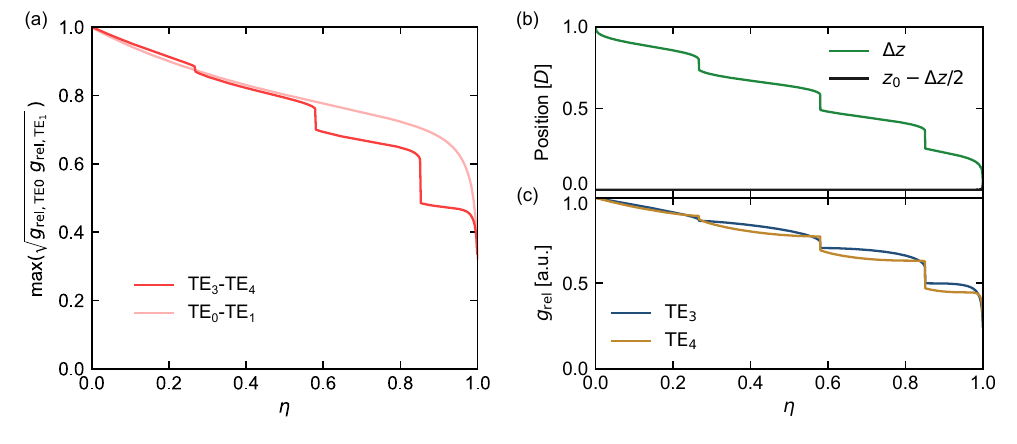}
	\caption{2D Optimization of the geometric mean of the coupling strengths for a given overlap $\eta_{45}$ using ideal TE$_3$ and TE$_4$ waveguide modes. (a)~Maximally achievable geometric mean of the coupling strengths for a given overlap. For comparison the previous result for coupling of TE$_0$ and TE$_1$ modes is shown in light red. (b)~The optimal values of the position and thickness of the active material to achieve the solution in (a).~The result of the optimization places the material at the bottom of the waveguide for all thicknesses. By symmetry, an equivalent solution exists at the top (position not shown for clarity). (c)~The resulting individual coupling strengths for the TE$_3$ and TE$_4$ modes for the solution in (a). $g_\mathrm{rel}=1$ corresponds to the coupling strength in a completely filled waveguide.}
	\label{FigSIdeal5}
\end{figure*}

By contrast, when the thickness is allowed to vary freely, as in the 2D optimization of the geometric mean of the coupling strengths at a fixed overlap (\cref{FigSIdeal5}), the optimal solution consistently places the active material at the top- or bottommost position. The results in~\cref{FigSIdeal5} now exhibit discontinuities at points where a further reduction in thickness no longer leads to an increase in overlap, thus requiring an abrupt change in the optimal thickness.

\clearpage

\begin{figure*}[ht] 
	\includegraphics[width=170mm]{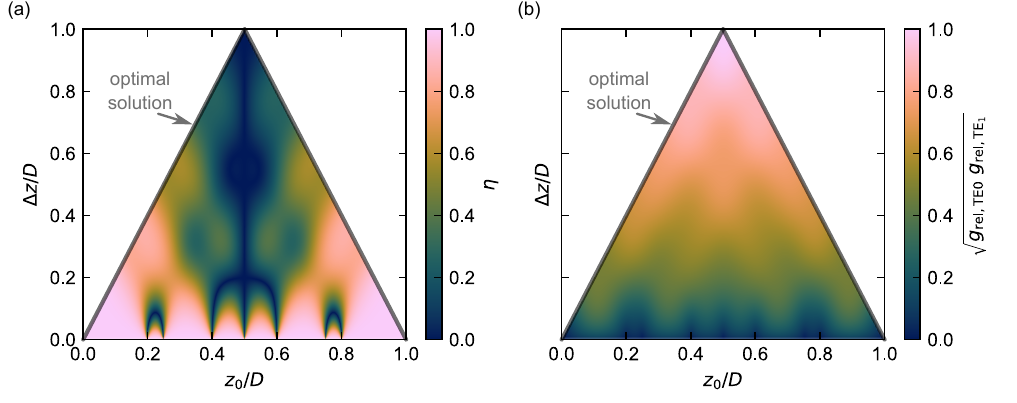}
	\caption{Dependence of (a)~the overlap $\eta_{45}$ and (b)~the geometric mean of the coupling strengths on the material thickness $\Delta z$ and position $z_0$ for ideal TE$_3$ and TE$_4$ waveguide modes. This library is used for the brute-force optimization shown in~\cref{FigSIdeal5}. The optimal solution, corresponding to placement of the active material at the top- or bottommost position of the waveguide, is highlighted by the gray line.}
	\label{FigSIdeal5b}
\end{figure*}

We note that this theoretical treatment of overlap and coupling strength applies both to waveguides and cavities as their idealized mode profiles are identical. The main difference lies in the fact that the energy spacing between modes can be made much smaller in waveguides compared to cavities. Therefore, superstrong coupling in cavities is often performed with higher-order modes, which results in a generally lower coupling strength (see comparison in~\cref{FigSIdeal5}(a)).

\section{Interference-based mode switching} \label{SI_Interference}

To gain physical insight into the interference-based switching scheme of~\cref{Fig6}, we derive an analytical expression for the wavevector shift of a polariton mode in response to a change in exciton energy. We consider a single photonic mode with energy $E_\mathrm{ph}(\beta)$ at a wavevector $\beta$, coupled to an exciton at energy $E_\mathrm{ex}$ with coupling strength $g$. The system is described by the $2 \times 2$ Hamiltonian:
\begin{equation}
	H = \begin{pmatrix} E_\mathrm{ph}(\beta) & g \\ g & E_\mathrm{ex} \end{pmatrix},
\end{equation}
with eigenvalues
\begin{equation} \label{Eq_Epm}
	E_j(\beta, E_\mathrm{ex}) = \frac{E_\mathrm{ph}(\beta) + E_\mathrm{ex}}{2} \pm \frac{1}{2}\sqrt{\delta_0^2 + 4g^2},
\end{equation}
where the sign is $-$ for $j=1$ (lower polariton) and $+$ for $j=2$ (upper polariton), and $\delta_0(\beta) = E_\mathrm{ph}(\beta) - E_\mathrm{ex}$ is the photon-exciton detuning. The exciton Hopfield coefficients of the two polariton branches are
\begin{equation} \label{Eq_Hopfield}
	|X_j|^2 = \frac{1}{2}\!\left(1 \mp \frac{\delta_0}{\sqrt{\delta_0^2 + 4g^2}}\right), \qquad |C_j|^2 = 1 - |X_j|^2,
\end{equation}
where the sign is $-$ for $j=1$ and $+$ for $j=2$.

In the switching scheme of~\cref{Fig6}, a laser at fixed energy $E_L$ excites a polariton at wavevector $\beta$ determined by the resonance condition $E_j(\beta, E_\mathrm{ex}) = E_L$. To find how $\beta$ responds to a shift in $E_\mathrm{ex}$ at constant $E_L$, we apply implicit differentiation:
\begin{equation}
	\frac{d\beta}{dE_\mathrm{ex}} = -\frac{\partial E_j / \partial E_\mathrm{ex}}{\partial E_j / \partial \beta}.
\end{equation}

Evaluating the partial derivatives of~\cref{Eq_Epm} yields:
\begin{align}
	\frac{\partial E_j}{\partial E_\mathrm{ex}} &= \frac{1}{2}\!\left(1 \mp \frac{\delta_0}{\sqrt{\delta_0^2 + 4g^2}}\right) = |X_j|^2, \label{Eq_dEdEx}\\[4pt]
	\frac{\partial E_j}{\partial \beta} &= \frac{\partial E_\mathrm{ph}}{\partial \beta}\;\frac{1}{2}\!\left(1 \pm \frac{\delta_0}{\sqrt{\delta_0^2 + 4g^2}}\right) = \hbar v_g^\mathrm{ph}\,|C_j|^2, \label{Eq_dEdbeta}
\end{align}
where $v_g^\mathrm{ph} = (1/\hbar)\,\partial E_\mathrm{ph}/\partial\beta$ is the group velocity of the uncoupled photonic mode. Equation~\eqref{Eq_dEdbeta} shows that the polariton group velocity is $v_{g,j} = v_g^\mathrm{ph}\,|C_j|^2$, i.e.\ the photon group velocity weighted by the photon fraction. Combining Eqs.~\eqref{Eq_dEdEx} and~\eqref{Eq_dEdbeta} gives the central result:
\begin{equation} \label{Eq_BetaSensitivity}
	\;\frac{d\beta}{dE_\mathrm{ex}} = -\frac{|X_j|^2}{\hbar\, v_{g,j}} = -\frac{|X_j|^2}{|C_j|^2}\;\frac{1}{\hbar\, v_g^\mathrm{ph}}.\;
\end{equation}

This expression has the following physical interpretation: polariton modes with a higher exciton fraction $|X_j|^2$ are more sensitive to changes in the exciton energy, while modes with a higher group velocity $v_{g,j}$ are less sensitive because a given energy shift translates into a smaller wavevector change along a steeper dispersion.

\begin{figure}[h]
    \centering
    \includegraphics[width=0.8\textwidth]{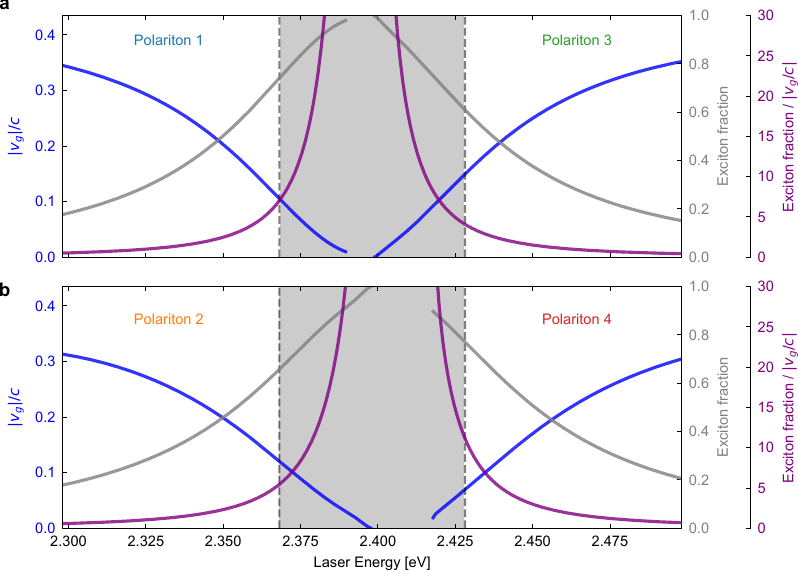}
	\caption{Parameters for the simplified $2\times2$ model for interference-based mode switching, using the dispersion calculated with~\cref{Hamiltonian} of the main text for the low-thickness waveguide (complementing~\cref{Fig6}). Exciton fraction (gray), group velocity (blue), and their ratio (purple) are shown for (a)~Polaritons~1 (left) and 3 (right), and (b)~Polaritons~2 (left) and 4 (right). The gray shaded area marks an energy range of $2\gamma_\mathrm{ex}$ centered at the exciton, where the system is too lossy for operation.}
    \label{FigS6}
\end{figure}

In the context of~\cref{Fig6}, the two co-propagating polaritons originate from different photonic modes ($\mathrm{TE}_0$ and $\mathrm{TE}_1$), each described by an independent $2 \times 2$ Hamiltonian with its own photon dispersion, coupling strength, and Hopfield coefficients. Applying~\cref{Eq_BetaSensitivity} to each branch, the figure of merit for the interference-based switching is thus:
\begin{equation} \label{Eq_FOM}
	\mathrm{FOM} = \left|\frac{d\beta_1}{dE_\mathrm{ex}} - \frac{d\beta_2}{dE_\mathrm{ex}}\right| = \left|\frac{|X_1|^2}{\hbar\, v_{g,1}} - \frac{|X_2|^2}{\hbar\, v_{g,2}}\right|,
\end{equation}
where the subscripts 1 and 2 refer to the two polariton modes. This FOM is maximized when the two polariton modes have strongly differing ratios of exciton fraction to group velocity, which occurs near the exciton resonance where the Hopfield coefficients change rapidly with the exciton energy (\cref{FigS6}).

\clearpage

\printbibliography